\documentclass[prl,aps,superscriptaddress,footinbib,10pt,twocolumn]{revtex4-2}
\usepackage{indentfirst}
\usepackage[utf8]{inputenc}
\usepackage{color}
\usepackage{amsmath}
\usepackage{amssymb}
\usepackage{stmaryrd}
\usepackage{graphicx}
\usepackage{siunitx}
\usepackage{scalerel}
\usepackage{adjustbox}
\usepackage{float}
\usepackage[unicode=true,
  bookmarks=true,bookmarksnumbered=false,bookmarksopen=false,
  breaklinks=false,pdfborder={0 0 1},backref=false,colorlinks=true]{hyperref}
\hypersetup{linkcolor=magenta, urlcolor=blue, citecolor=blue, pdfstartview={FitH}, hyperfootnotes=false, unicode=true}
\makeatletter

\usepackage{amsfonts}
\usepackage{tabularx}
\usepackage{dcolumn}
\usepackage{bm}
\usepackage{graphicx}
\usepackage{epstopdf}
\usepackage{times}
\hypersetup{urlcolor=blue}

\def\ket#1{\left|#1\right\rangle}
\def\bra#1{\left\langle#1\right|}

\makeatother

\begin{document}

\title{Real-time scattering and freeze-out dynamics in Rydberg-atom lattice gauge theory}

\author{De-Sheng Xiang}
\thanks{These authors contributed equally to this work.}
\affiliation{MOE Key Laboratory of Fundamental Physical Quantities Measurement,
Hubei Key Laboratory of Gravitation and Quantum Physics, PGMF,
Institute for Quantum Science and Engineering, School of Physics,
Huazhong University of Science and Technology, Wuhan 430074, China}

\author{Peng Zhou}
\thanks{These authors contributed equally to this work.}
\affiliation{MOE Key Laboratory of Fundamental Physical Quantities Measurement,
Hubei Key Laboratory of Gravitation and Quantum Physics, PGMF,
Institute for Quantum Science and Engineering, School of Physics,
Huazhong University of Science and Technology, Wuhan 430074, China}

\author{Chang Liu}
\thanks{These authors contributed equally to this work.}
\affiliation{Shanghai Qi Zhi Institute,  Shanghai 200232, China}

\author{Hao-Xiang Liu}
\affiliation{MOE Key Laboratory of Fundamental Physical Quantities Measurement,
Hubei Key Laboratory of Gravitation and Quantum Physics, PGMF,
Institute for Quantum Science and Engineering, School of Physics,
Huazhong University of Science and Technology, Wuhan 430074, China}

\author{Yao-Wen Zhang}
\affiliation{MOE Key Laboratory of Fundamental Physical Quantities Measurement,
Hubei Key Laboratory of Gravitation and Quantum Physics, PGMF,
Institute for Quantum Science and Engineering, School of Physics,
Huazhong University of Science and Technology, Wuhan 430074, China}

\author{Dong Yuan}
\affiliation{Center for Quantum Information, IIIS, Tsinghua University, Beijing 100084, China}
\affiliation{JILA, University of Colorado Boulder, Boulder, Colorado 80309, USA}

\author{Kuan Zhang}
\affiliation{MOE Key Laboratory of Fundamental Physical Quantities Measurement,
Hubei Key Laboratory of Gravitation and Quantum Physics, PGMF,
Institute for Quantum Science and Engineering, School of Physics,
Huazhong University of Science and Technology, Wuhan 430074, China}

\author{Biao Xu}
\affiliation{MOE Key Laboratory of Fundamental Physical Quantities Measurement,
Hubei Key Laboratory of Gravitation and Quantum Physics, PGMF,
Institute for Quantum Science and Engineering, School of Physics,
Huazhong University of Science and Technology, Wuhan 430074, China}

\author{Marcello Dalmonte}
\email{mdalmont@ictp.it}
\affiliation{The Abdus Salam International Centre for Theoretical Physics (ICTP), 
Strada Costiera 11, Trieste 34151, Italy}

\author{Dong-Ling Deng}
\email{dldeng@tsinghua.edu.cn}
\affiliation{Center for Quantum Information, IIIS, Tsinghua University, Beijing 100084, China}
\affiliation{Shanghai Qi Zhi Institute,  Shanghai 200232, China}
\affiliation{Hefei National Laboratory, Hefei 230088, China}

\author{Lin Li}
\email{li\_lin@hust.edu.cn}
\affiliation{MOE Key Laboratory of Fundamental Physical Quantities Measurement,
Hubei Key Laboratory of Gravitation and Quantum Physics, PGMF,
Institute for Quantum Science and Engineering, School of Physics,
Huazhong University of Science and Technology, Wuhan 430074, China}
\affiliation{Wuhan Institute of Quantum Technology, Wuhan 430206, China}

\begin{abstract}
{Understanding the non-equilibrium dynamics of gauge theories remains a fundamental challenge in high-energy physics ~\cite{Calzetta_Hu_2008,  gross_50_2023}. Indeed, most large scale experiments on gauge theories intrinsically rely on very far-from equilibrium dynamics, from heavy-ion to lepton and hadron collisions, which is in general extremely challenging to treat ab initio ~\cite{bauer_quantum_2023-1}. Quantum simulation holds intriguing potential in tackling this problem \cite{jordan_quantum_2012, zohar_quantum_2016, Bañuls2020, Karpov_Spatiotemporal_2022, bauer_quantum_2023, bauer_quantum_2023-1, di_meglio_quantum_2024, su_cold-atom_2024, cheng_emergent_2024, halimeh_cold-atom_2025, farrell_digital_2025} and pioneering experiments have observed different characteristic features of gauge theories \cite{martinez_real-time_2016,kokail_self-verifying_2019, tan_observation_2021, meth_simulating_2025, mueller_quantum_2025,schweizer_floquet_2019, gorg_realization_2019, yang_observation_2020, mil_scalable_2020, frolian_realizing_2022,  zhou_thermalization_2022, wang_interrelated_2023, zhang_observation_2025,wang_observation_2022,mildenberger_confinement_2025,cobos2025realtimedynamics21dgauge,bernien2017probing,surace_lattice_2020,  datla_statistical_2025,gonzalez-cuadra_observation_2025, Cochran2025,de_observation_2024, liu_string_2024,zenesini_false_2024,zhu_probing_2024, vodeb_stirring_2025}, such as string breaking \cite{gonzalez-cuadra_observation_2025, Cochran2025,de_observation_2024, liu_string_2024} and false vacuum decay~\cite{zenesini_false_2024, zhu_probing_2024, vodeb_stirring_2025}. Here, using a programmable Rydberg atom array, we observe real-time scattering and freeze-out dynamics in a (1+1)-dimensional U(1) lattice gauge theory.  
Through spatiotemporal Hamiltonian engineering, we demonstrate dynamical confinement-deconfinement transitions, revealing string fragmentation and symmetry restoration during quenches. We track scattering processes with single-site resolution across a range of parameter regimes. Utilizing a double quench protocol, we observe dynamical freeze-out: upon quenching the Hamiltonian after scattering, despite the injection of an extensive energy, the system evolution---in terms of both low-order correlations and entanglement---freezes, effectively stabilizing a highly correlated equilibrium state---a situation that reminisces that of collisions between heavy ions~\cite{Berges_QCD_2021}. Our work establishes a high-resolution approach for probing non-perturbative gauge dynamics~\cite{CarmenBañuls_Review_2020}, opening alternative pathways toward studying far-from-equilibrium phenomena in high-energy physics.}

\end{abstract}

\maketitle

\noindent  Gauge theory governs three fundamental interactions, with quantum chromodynamics (QCD) ~\cite{gross_50_2023}—the theory of the strong interaction—exhibiting profound phenomena such as confinement ~\cite{wilson_confinement_1974}, string breaking ~\cite{bali_observation_2005}, and asymptotic freedom ~\cite{Frank_Nobel_2005}. Theoretically, QCD is described by a relatively simple set of equations based on gauge theory. Yet, solving these equations in an analytical fashion pose notorious challenges due to strong interactions between quarks and gluons~\cite{wiese_ultracold_2013}. On the numerical side, while lattice gauge theory (LGT) ~\cite{wilson_confinement_1974,creutz_monte_1983}  provides a powerful non-perturbative framework for equilibrium properties,  from the low-lying spectrum to the phase diagram, its resource demands scale exponentially for real-time dynamics ~\cite{troyer_computational_2005, CarmenBañuls_Review_2020}. Experimentally, particle colliders detect asymptotic final states but lack the controllability to probe transient intermediates~\cite{belyansky_high-energy_2024}. This constraint is particularly acute for QCD scattering processes, often evolving at ultrafast timescales, where evolution cannot be paused and intermediate stages evade direct observation.  These theoretical and experimental challenges make far-from-equilibrium gauge dynamics a central frontier in fundamental physics.

\begin{figure*}[t]
  \centering
  \includegraphics[width=\textwidth]{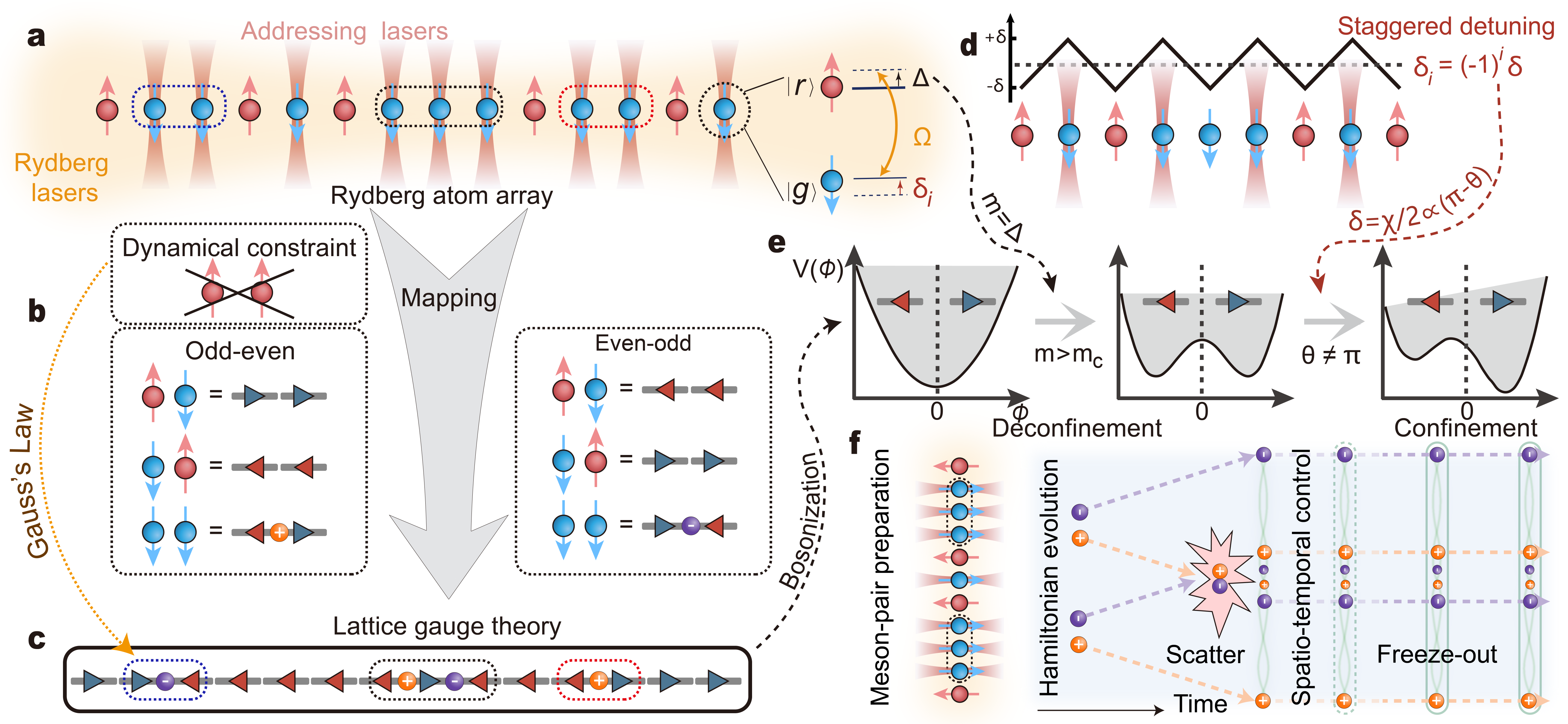}
\caption{
\textbf{Rydberg lattice gauge theory quantum simulator.} 
\textbf{a}, Programmable Rydberg-atom array. Atoms are globally driven by Rydberg lasers with effective Rabi frequency $\Omega$ and detuning $\Delta$, and individually addressed via far-detuned lasers which provide local light shifts $\delta_i$ (Methods).
\textbf{b}, Lattice gauge theory (LGT) mapping. Atomic states $\ket{g}$ and $\ket{r}$ encode gauge fields $+E$ (blue arrow) and $-E$ (red arrow) on odd sites, with the atom-field correspondence inverted on even sites. The Rydberg blockade effect enforces Gauss's law.  At matter sites (defined as the interstitial positions between adjacent gauge sites), positive and negative charges emerge as domain walls between opposite gauge polarities.
\textbf{c}, Realization of the quantum link model. The LGT configuration directly maps to the atomic state arrangement in the array shown in \textbf{a}, exhibiting alternating gauge fields with localized matter excitations.
\textbf{d}, Control of the LGT Hamiltonian. Global detuning $\Delta$ modulates fermion mass $m$ from $0$ to $m>m_c$, sculpting $V(\phi)$ between single-minimum and double-well structures in \textbf{e}. Staggered detuning $\delta_i = (-1)^i\delta$ shifts $\theta$ away from $\pi$, driving confinement-deconfinement transitions via potential asymmetry.
\textbf{e}, Confinement-deconfinement mechanism. At $\theta=\pi$ (deconfinement), degenerate minima of $V(\phi)$ enable free propagation with vanishing string tension. For $\theta \neq \pi$ (confinement), asymmetry induces non-zero string tension, imposing an energy cost that increases linearly with separation distance, suppressing pair creation. 
\textbf{f}, Dynamical control of scattering dynamics. High-fidelity meson-like initial state is prepared through combined global Rydberg Raman $\pi$-pulse and large-scale local shelving beams (Methods). Subsequently, spatio-temporal control of the LGT Hamiltonian parameters during particle scattering triggers quantum freeze-out dynamics and enables the freezing of intermediate scattering states.
}
  \label{Fig1:experimental_set_up}
\end{figure*}

Quantum simulation offers a resource-efficient alternative to colliders and classical computations ~\cite{jordan_quantum_2012, zohar_quantum_2016, Bañuls2020, Karpov_Spatiotemporal_2022, bauer_quantum_2023, bauer_quantum_2023-1, di_meglio_quantum_2024, su_cold-atom_2024, cheng_emergent_2024, halimeh_cold-atom_2025, farrell_digital_2025}, enabling table-top studies of LGTs through platforms such as trapped ions ~\cite{martinez_real-time_2016,kokail_self-verifying_2019, tan_observation_2021, meth_simulating_2025, mueller_quantum_2025}, superconducting qubits ~\cite{wang_observation_2022,Cochran2025,mildenberger_confinement_2025,cobos2025realtimedynamics21dgauge}, ultracold atoms in optical lattices ~\cite{schweizer_floquet_2019, gorg_realization_2019, yang_observation_2020, mil_scalable_2020, frolian_realizing_2022,  zhou_thermalization_2022, wang_interrelated_2023, zhang_observation_2025}, and Rydberg atom arrays ~\cite{bernien2017probing,surace_lattice_2020, gonzalez-cuadra_observation_2025, datla_statistical_2025}. These systems engineer Hamiltonians that approximate LGTs, replicating key non-perturbative features including confinement ~\cite{schweizer_floquet_2019, tan_observation_2021, mildenberger_confinement_2025, zhang_observation_2025} and nontrivial vacuum structure ~\cite{zhu_probing_2024}, thus providing intriguing testbeds for  intricate QCD dynamics. However, despite the fact that exciting progresses have been made along the direction of simulating LGTs with various platforms \cite{martinez_real-time_2016,kokail_self-verifying_2019, tan_observation_2021, meth_simulating_2025, mueller_quantum_2025,schweizer_floquet_2019, gorg_realization_2019, yang_observation_2020, mil_scalable_2020, frolian_realizing_2022,  zhou_thermalization_2022, wang_interrelated_2023, zhang_observation_2025,wang_observation_2022,mildenberger_confinement_2025,cobos2025realtimedynamics21dgauge,bernien2017probing,surace_lattice_2020,  datla_statistical_2025,gonzalez-cuadra_observation_2025, Cochran2025,de_observation_2024, liu_string_2024,zenesini_false_2024,zhu_probing_2024, vodeb_stirring_2025}, resolving large-scale non-equilibrium dynamics remains challenging due to limitations in state preparation, spatiotemporal Hamiltonian engineering, and high-resolution probing ~\cite{su_cold-atom_2024}. Overcoming these hurdles is essential to explore fundamental phenomena—from recently observed false vacuum decay ~\cite{zenesini_false_2024, zhu_probing_2024, vodeb_stirring_2025} and string breaking \cite{de_observation_2024, liu_string_2024, gonzalez-cuadra_observation_2025, Cochran2025} to the uncharted territory of real-time particle scattering \cite{su_cold-atom_2024, belyansky_high-energy_2024}.

Here, we bridge this critical gap by leveraging a programmable Rydberg quantum simulator with spatiotemporally tunable Hamiltonian control to track and tailor the full real-time dynamics of a $U(1)$ LGT. Through precise dynamical engineering of the topological $\theta$-angle and fermion mass, we demonstrate quantum quench-induced confinement-deconfinement transitions characterized by string fragmentation and charge-parity symmetry breaking.
We further probe real-time charge scattering dynamics in the deconfined regime, and observe diamond-shaped interference patterns, signifying (1+1)D quasi-elastic scattering. 
Leveraging Hamiltonian spatiotemporal control, we change the Hamiltonian dynamics exactly at collision onset. Despite the highly non-equilibrium nature of the process, we observe a clear dynamical freeze-out---indicating that the highly energetic scattering state generated by the dynamics approximately describes an equilibrium state of the quenched Hamiltonian. This observation is reminiscent of freeze-out descriptions of heavy ion collisions~\cite{borsanyi2013freeze, Heinz:2007in, PhysRevLett.109.192302}, where the result of a highly non-perturbative, non-equilibrium process admits an effective equilibrium description.

\begin{figure} [t]
  \centering
  \includegraphics[width=\columnwidth]{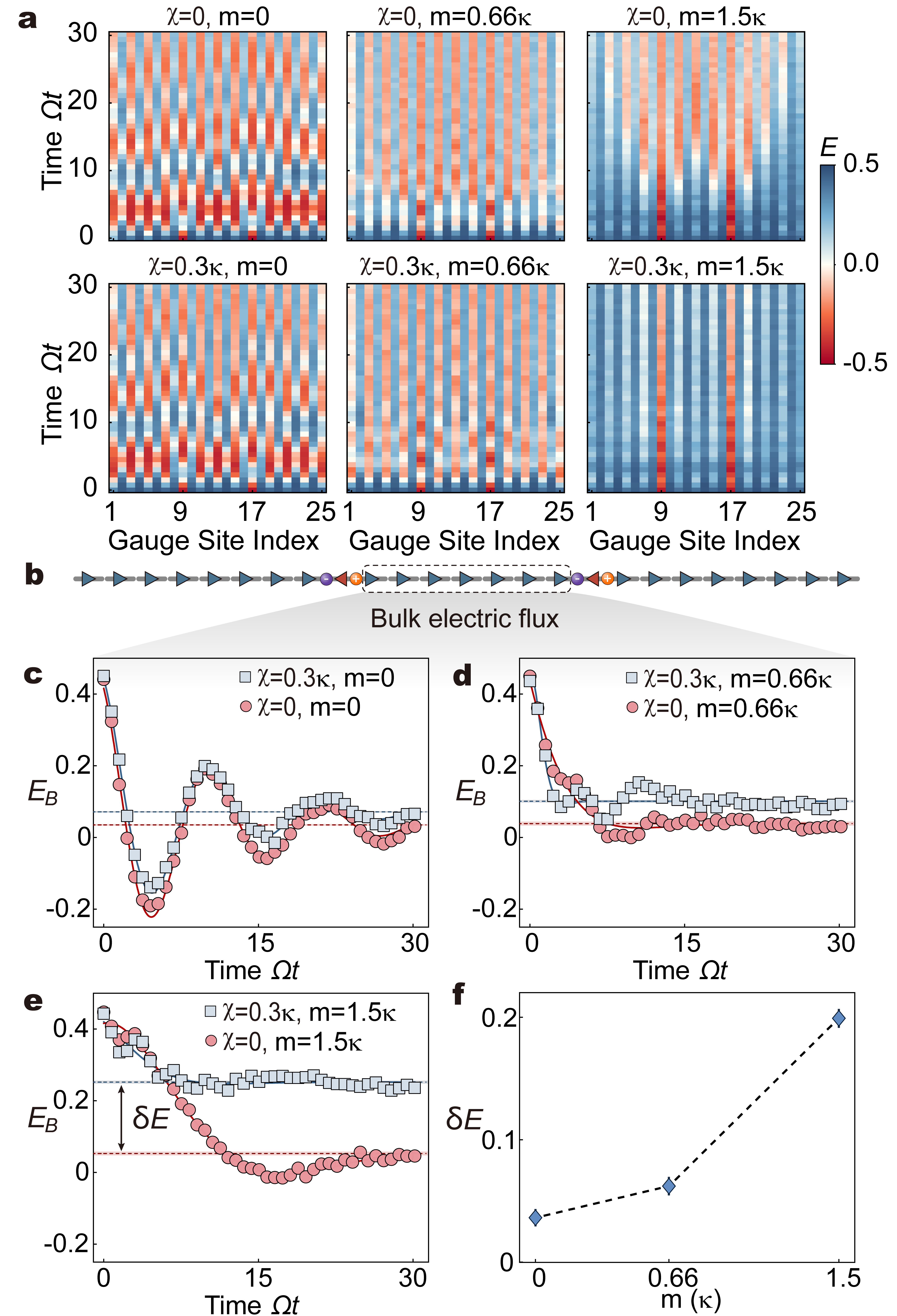}
  \caption{
    \textbf{Confinement and deconfinement.}
\textbf{a,} Spatiotemporal evolution of the electric field $E_i(t)$ for different fermion mass $m$ and topological $\theta$ angle term (quantified by string tension $\chi$). \textbf{b,} Electric field and charge configuration of the tailored initial state.
\textbf{c--e,} Evolution of the averaged bulk electric flux $E_\mathrm{B}(t)$ under different mass values. 
Red circles: deconfinement regime ($\theta = \pi$,  corresponding to $\chi = 0$);
blue squares: confinement phase ($\theta \neq \pi$, $\chi = 0.3\kappa$). 
Solid lines: damped oscillatory fits to the experimental data (Methods). Dashed lines: steady-state values (error bars denoted by the shaded bands).
In the deconfinement phase, the bulk electric flux stabilizes near zero (red dashed line). In the confinement regime, it converges to a shifted steady-state value (blue dashed line), reflecting the asymmetric potential $V(\phi)$ at $\theta \neq \pi$.
\textbf{f,} Steady-state bias $\delta E$ of the bulk electric flux as a function of mass. Error bars represent 68\% confidence intervals.  
}
  \label{Fig:2_confinement_and_deconfinement_phases}
\end{figure}

\vspace{.5cm}

\noindent\textbf{\large{}The model and experimental setup}

\noindent Our experimental platform consists of a programmable array of up to 30 $^{87}$Rb atoms (Fig.~\ref{Fig1:experimental_set_up}\textbf{a}). The atoms are initially prepared in the ground state $|g\rangle$ and coupled to a Rydberg state $|r\rangle$ via a two-photon transition with effective Rabi frequency $\Omega$ and global detuning $\Delta$.
Site-selective control is achieved using far-detuned addressing lasers, which program local AC-Stark shifts $\delta_i$ at each atomic site. This capability enables tailored detuning patterns—including the staggered configuration $\delta_i = (-1)^i\delta$ for tuning the topological $\theta$-angle (Fig.~\ref{Fig1:experimental_set_up}\textbf{d}-\textbf{e}).
The dynamics of the Rydberg atom array is described by the following Hamiltonian:
$
\hat{H}_{R}/\hbar = \sum_i \left[ \frac{\Omega}{2} \hat{\sigma}_i^x - (\Delta + \delta_i) \hat{n}_i \right] + \sum_{i<j} V_{ij} \hat{n}_i \hat{n}_j,
$
where $\hat{\sigma}_i^x$ represents the coherent coupling between ground and Rydberg states, $\hat{n}_i$ is the Rydberg state occupation operator, $V_{ij}$ describes the van der Waals interactions between atoms.

In the Rydberg blockade regime, i.e. $V_{i,i+1} \gg \Omega$, adjacent atoms cannot be simultaneously excited to Rydberg states. Under this dynamical constraint, $\hat{H}_{R}$ can be mapped onto the Hamiltonian of a U(1) LGT~\cite{ surace_lattice_2020} in the quantum link model formulation~\cite{banerjee_atomic_2012} (Methods):
\begin{multline}
\hat{H}_{\text{LGT}} = -\kappa \sum_i \left( \hat{\psi}_i^\dagger \hat{U}_{i,i+1} \hat{\psi}_{i+1} + \text{H.c.} \right) \\
+ m \sum_i (-1)^i \hat{\psi}_i^\dagger \hat{\psi}_i
+ J \sum_i \left( \hat{E}_{i,i+1} + \frac{\theta}{2\pi} \right)^2.
\end{multline}
Here, $\hat{\psi}_i$ ($\hat{\psi}_i^\dagger$) represents the annihilation (creation) operators for staggered fermion at site $i$. $\hat{U}_{i,i+1}$ denotes the dynamical gauge field at the link between sites $i$ and $i+1$, and is represented as a spin-1/2 annihilation operator.
Its conjugate is the electric field $\hat{E}_{i,i+1}$, and satisfies the commutation relation $[\hat{E}_{i,i+1}, \hat{U}_{i,i+1}] = \hat{U}_{i,i+1}$. The LGT dynamics is described by a gauge-matter coupling $\kappa$, the staggered fermion mass $m$, and the gauge-invariant electric field energy $J$.
In our experiment, these terms can be independently controlled through the Rabi frequency $\Omega = -\kappa$, the global detuning $\Delta= m$, and the staggered detuning $2\delta = \chi = J(\pi - \theta)/\pi$, where $\chi$ is the effective string tension (Methods).

In the LGT model, the electric field $\hat{E}_{i,i+1}=(-1)^i\hat{\sigma}_i^z$ can be interpreted as arising from domain wall configurations between excited and unexcited regions.
The dynamical charge at matter site $i$ is defined as $\hat{Q}_i = \hat{\psi}_i^\dagger\hat{\psi}_i - (1+(-1)^i)/2$.
The Rydberg blockade effect enforces Gauss's law~\cite{surace_lattice_2020}, $G_i = E_{i,i+1} - E_{i-1,i} - Q_i = 0$, ensuring local charge conservation. This law manifests as gauge invariance in the LGT model, a core feature of the theory.
The correspondence between dynamical charges, surrounding electric fields, and atomic state configurations is illustrated in Fig.~\ref{Fig1:experimental_set_up}\textbf{b}-\textbf{c}.
In the continuum limit, the LGT Hamiltonian can be bosonized and mapped onto a sine-Gordon model with a scalar potential $V(\phi) = e^2\phi^2 - cm \cos(2\phi - \theta)$ 
(Methods)~\cite{COLEMAN1976239,surace_lattice_2020}.
Here, $e$ is the gauge coupling constant, $c$ is a constant determined by $\Omega$ and $\Delta$, and $\phi$ is the bosonized scalar field. 
The scalar potential $V(\phi)$ determines the vacuum structure~\cite{CALLAN1976334} and dynamical behavior of the system (Fig.~\ref{Fig1:experimental_set_up}\textbf{e}), leading to fundamental phenomena including confinement-deconfinement transitions~\cite{surace_lattice_2020} and spontaneous charge-parity symmetry breaking~\cite{yang_observation_2020}.

Experimentally, by precisely engineering the interplay between global Rydberg driving and site-resolved addressing lasers (Methods), we can prepare high-fidelity atomic configurations that encode various LGT states, including vacua with uniform positive or negative electric fields, isolated charges, and meson-like excitations (charged particle-antiparticle pairs).
Combining high-fidelity initial state preparation with precise control of $V(\phi)$ enables direct access to real-time scattering dynamics (Fig.~\ref{Fig1:experimental_set_up}\textbf{f}).
By preparing spatially separated mesons and quenching to deconfinement ($\theta=\pi$), we trigger charge propagation and collisions. 
Furthermore, the spatiotemporal programmability of the LGT Hamiltonian allows abrupt confinement quenches during scattering, leading to the observation of the emergent quantum freeze-out phenomenon (Fig.~\ref{Fig1:experimental_set_up}\textbf{f}).

\vspace{.5cm}

\begin{figure*}[t]
  \centering
  \includegraphics[width=\textwidth]{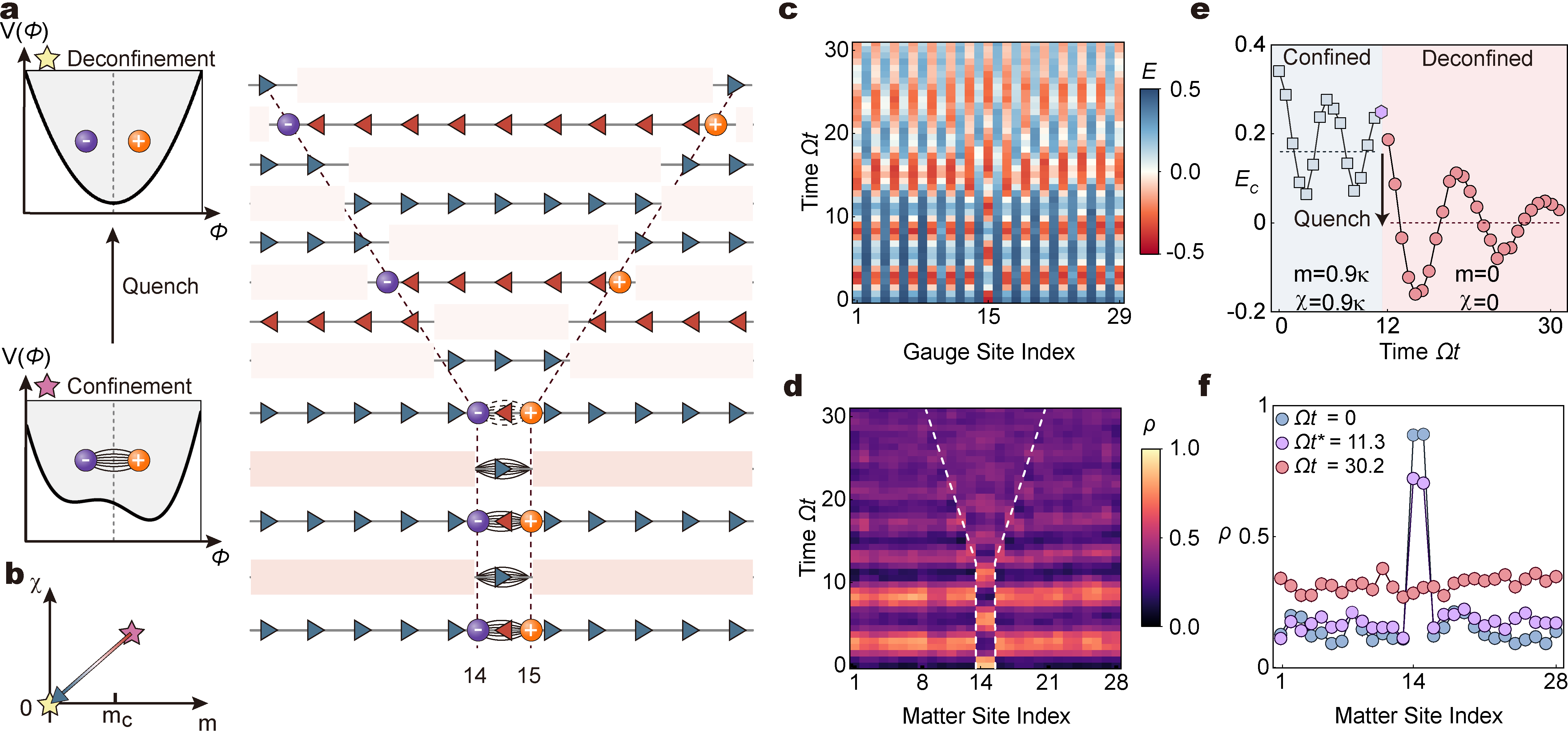}
  \caption{
    \textbf{Quench dynamics.} 
\textbf{a}, Schematic illustration of dual-parameter quench protocol. In the confinement regime, initially prepared negative and positive charges are connected by a negative electric flux tube (red arrows) and bound by string tension $\chi=1.8\kappa$. 
At $\Omega t^*=11$, rapid dual-parameter quench ($\theta \rightarrow \pi$, $m \rightarrow 0$) to the deconfinement phase triggers string fragmentation, pair production, and light-cone charge propagation.
Shaded regions: high charge density with vanishing  average electric field $\langle E_{j,j+1}\rangle\simeq0$. 
Insets: effective potentials $V(\phi)$ in the confined and deconfined phases.
\textbf{b}, Experimental quench trajectory shown in the $\chi$--$m$ phase diagram.
Red (yellow) star indicates the pre- (post-) quench parameter.
\textbf{c--d}, Measured spatiotemporal evolution of the electric field $E_i(t)$ (\textbf{c}) and charge density $\rho_i(t)$ (\textbf{d}).
\textbf{e}, Dynamics of the averaged central electric flux $E_c(t)$. 
Blue squares (red circles): $E_c(t)$ in the confined (deconfined) phase.
Purple hexagon: the quench point.
Dashed lines: oscillatory baselines. 
The arrow indicates quench operation.
\textbf{f}, Spatial charge distributions at key times: pre-quench ($\Omega t=0$, blue), quench point ($\Omega t^*=11.3$, purple), and post-quench ($\Omega t=30.2$, red). 
}
  \label{Fig3:Quench_dynamics}
\end{figure*}

\vspace{.5cm}

\noindent\textbf{\large{}Confinement-deconfinement transition}

\noindent  
The characterization of confinement and deconfinement phases establishes the essential framework for probing real-time scattering and freeze-out dynamics. 
To this end, we employ the effective string tension $\chi$, which quantifies the linear energy cost for separating opposite charges in the static limit (Methods).
Experimentally, we control $\chi$ by simultaneously tuning the fermion mass $m$ and topological $\theta$-angle, while monitoring the evolution dynamics of the single-site resolved electric field $E_i(t)$ and the averaged bulk electric flux $E_\mathrm{B}(t) = \frac{1}{7}\sum_{i=10}^{16} E_i(t)$. The initial state encodes two meson-like excitations at sites $9$ and $17$ via the electric field configuration $E_i(0)=0.5-\delta_{i,9}-\delta_{i,17}$, embedded within the vacuum background (Fig.~\ref{Fig:2_confinement_and_deconfinement_phases}\textbf{b}).

In the deconfined regime ($\theta=\pi$), the scalar potential $V(\phi)$ exhibits distinct vacuum structures (Fig.~\ref{Fig1:experimental_set_up}\textbf{e}). At $m=0$, it possesses a single minimum at $\phi=0$, corresponding to a charge-parity symmetric vacuum with oscillating and vanishing electric field (Fig.~\ref{Fig:2_confinement_and_deconfinement_phases}\textbf{b}, red circles). When the fermion mass exceeds the critical value $m_c$, degenerate minima emerge at $\phi=\pm\phi_0$, resulting in spontaneous charge-parity symmetry breaking and long-time relaxation of the electric field to a stable minimum (Fig.~\ref{Fig:2_confinement_and_deconfinement_phases}\textbf{e}, red circles). This vacuum degeneracy eliminates the energy barrier for particle-antiparticle pair creation, yielding vanishing string tension ($\chi=0$) and unimpeded charge propagation. Our measurements directly capture these deconfinement dynamics through rapid dispersion of the initial electric flux from sites 9 and 17 throughout the system (Fig.~\ref{Fig:2_confinement_and_deconfinement_phases}\textbf{a}, upper panel).

Confinement emerges when topological $\theta$ angle deviates from $\pi$, experimentally implemented via staggered detuning $\delta=0.15\Omega$ (corresponding to $\chi=0.3\kappa$). This lifts the vacuum degeneracy, establishing a gap proportional to $\chi$ between opposite electric field orientations (Fig.~\ref{Fig1:experimental_set_up}\textbf{e}) and generating the characteristic linear confining potential. The observed confinement dynamics display strong mass dependence: at $m=0$, the initial electric flux disperses rapidly due to negligible confinement, while increasing mass enhances charge localization (Fig.~\ref{Fig:2_confinement_and_deconfinement_phases}\textbf{a}, lower panel). Furthermore, the bulk electric flux develops a mass-dependent steady-state bias  $\delta E$ (Fig.~\ref{Fig:2_confinement_and_deconfinement_phases}\textbf{c}-\textbf{e}), reflecting enhanced asymmetry in $V(\phi)$ (Methods) and strengthened confinement at larger mass. Complementing these findings, we also demonstrate that increasing string tension $\chi$ leads to stronger confinement (Methods).

\vspace{.5cm}

\noindent\textbf{\large{}Dynamical control of LGT Hamiltonian}

\noindent 
Capturing scattering intermediates demands spatiotemporal Hamiltonian control. Building on the characterized static properties of the confinement and deconfinement phases, we therefore engineer rapid quench protocols—leveraging precise parameter modulation—to achieve in situ phase switching within a single experimental trial. 
Beyond enabling real-time studies of string fragmentation and symmetry restoration, this dynamical capability allows for the direct observation of freeze-out dynamics.

The quench experiment begins with the preparation of a mesonic excitation at site 15, encoded through the electric field configuration $E_i(0) = 0.5 - \delta_{i,15}$ within the vacuum background.
With staggered detuning $\delta= 0.9 \Omega$ (i.e.~$\theta \neq \pi$) and mass $m = 0.9 \kappa$, the system initially resides in the confinement phase where the charges remain bound through stable electric-flux configurations due to the vacuum structure of $V(\phi)$ with non-zero string tension. 
At $\Omega t^* = 11.3$, a dual-parameter quench executed within 20 ns (much shorter than the experimental time
step, Methods) instantaneously sets $m = 0$ and $\theta = \pi$. This drives the system into the deconfined regime where vacuum degeneracy emerges in $V(\phi)$, eliminating the energy barrier for pair creation (Fig.~\ref{Fig3:Quench_dynamics}\textbf{a}). 
The vanishing string tension accompanies charge-parity symmetry restoration, triggering out-of-equilibrium dynamics characterized by string fragmentation and proliferating pair production. Spatiotemporal measurements of electric field $E_i(t)$ and charge density $\rho_i(t) = |Q_i(t)|$ (Methods) reveal a sharp dynamical transformation: pre-quench oscillations arising from Hilbert space fragmentation at the $\delta = \Delta$ resonance~\cite{desaules_ergodicity_2024, Variational_shi.T_2018} are abruptly replaced by post-quench light-cone propagation (Fig.~\ref{Fig3:Quench_dynamics}\textbf{d}).

This quench-induced transition is further characterized by employing the averaged central electric flux $E_c(t) = \frac{1}{14}\sum_{i=8}^{21} E_i(t)$  as an order parameter. 
The measured $E_c(t)$ dynamics (Fig.~\ref{Fig3:Quench_dynamics}\textbf{e}) exhibit a discontinuous shift in the oscillatory baseline (denoted by dashed lines) at the quench point, signifying a transition between confinement and deconfinement. 
Measured spatial charge distributions in Fig.~\ref{Fig3:Quench_dynamics}\textbf{f} also corroborate this transition, with particles evolving from confinement ($\Omega t=0$) to delocalization ($\Omega t=30.2$) within a single coherent evolution enabled by quantum quench dynamics.

\vspace{.5cm}
\noindent\textbf{\large{}Scattering and freeze-out dynamics}

\noindent High-energy scattering of composite particles like mesons provides fundamental probes of nonperturbative gauge dynamics, exhibiting hallmark phenomena including confinement and bound-state formation or dissociation through complex multistage processes~\cite{belyansky_high-energy_2024}. Despite advances in nonperturbative theoretical methods, experimental studies of real-time scattering dynamics remain elusive due to challenges in preparing controlled initial states and resolving ultrafast processes. Rydberg quantum simulators for lattice gauge theories now offer new pathways to investigate these dynamics~\cite{Karpov_Spatiotemporal_2022, bauer_quantum_2023-1}. 
Leveraging our Rydberg atom array with site-and-time-resolved control, we investigate scattering and freeze-out dynamics by initializing two meson-like excitations separated by seven sites in a vacuum state, with fermion mass $m = 1.5\kappa$  to suppress vacuum fluctuations. Differently from other numerically investigated settings~\cite{pichler_real-time_2016,bauer_quantum_2023-1}, here the scattering we are interested in is that between the quasiparticles generated after quenching a meson state---that is, those investigated in the previous section. This setup has the advantage that there is no need of specific wave-packet initialization, and is robust to imperfections as long as those do not affect meson melting.  

\begin{figure}[t!]
  \includegraphics[width=\columnwidth]{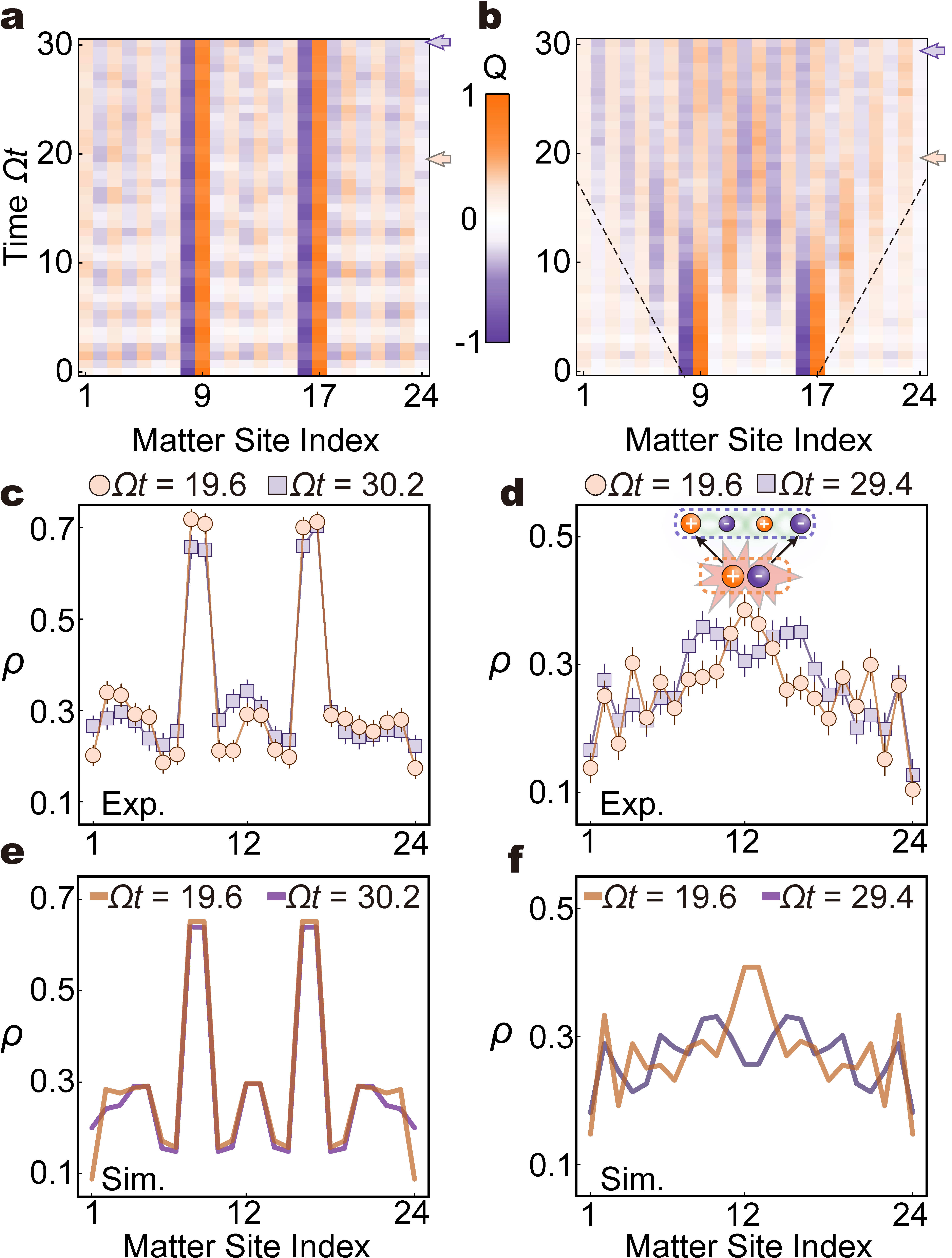}
  \caption{\textbf{Real-time scattering dynamics in the U(1) lattice gauge theory.} 
  \textbf{a, b}, Experimentally measured spatiotemporal evolution of the dynamical charge $Q_i(t)$ under varying string tensions. \textbf{c, d}, Corresponding measured site-resolved particle density $\rho_i(t) = |Q_i(t)|$ at selected evolution times. \textbf{a, c}, In the strong confinement regime ($\chi=0.6\kappa$), charges remain localized with density concentrated at initial positions throughout evolution. \textbf{b, d}, In the deconfinement phase ($\chi=0$), charges propagate freely and scatter when their wave packet overlap. A central density peak is observed at $\Omega t=19.6$, which evolves into a symmetric double-peak structure at $\Omega t=29.4$. Dashed lines in \textbf{b} indicate the light-cone trajectories. Inset of \textbf{d} shows a schematic illustration of the collision. \textbf{e, f}, Numerical simulations of the corresponding site-resolved particle density  for the confinement (\textbf{e}) and deconfinement (\textbf{f}) regimes. Error bars represent one standard deviation.
}
  \label{Fig:4_Meson_dynamics}
\end{figure}

\begin{figure}[t!]
  \includegraphics[width=\columnwidth]{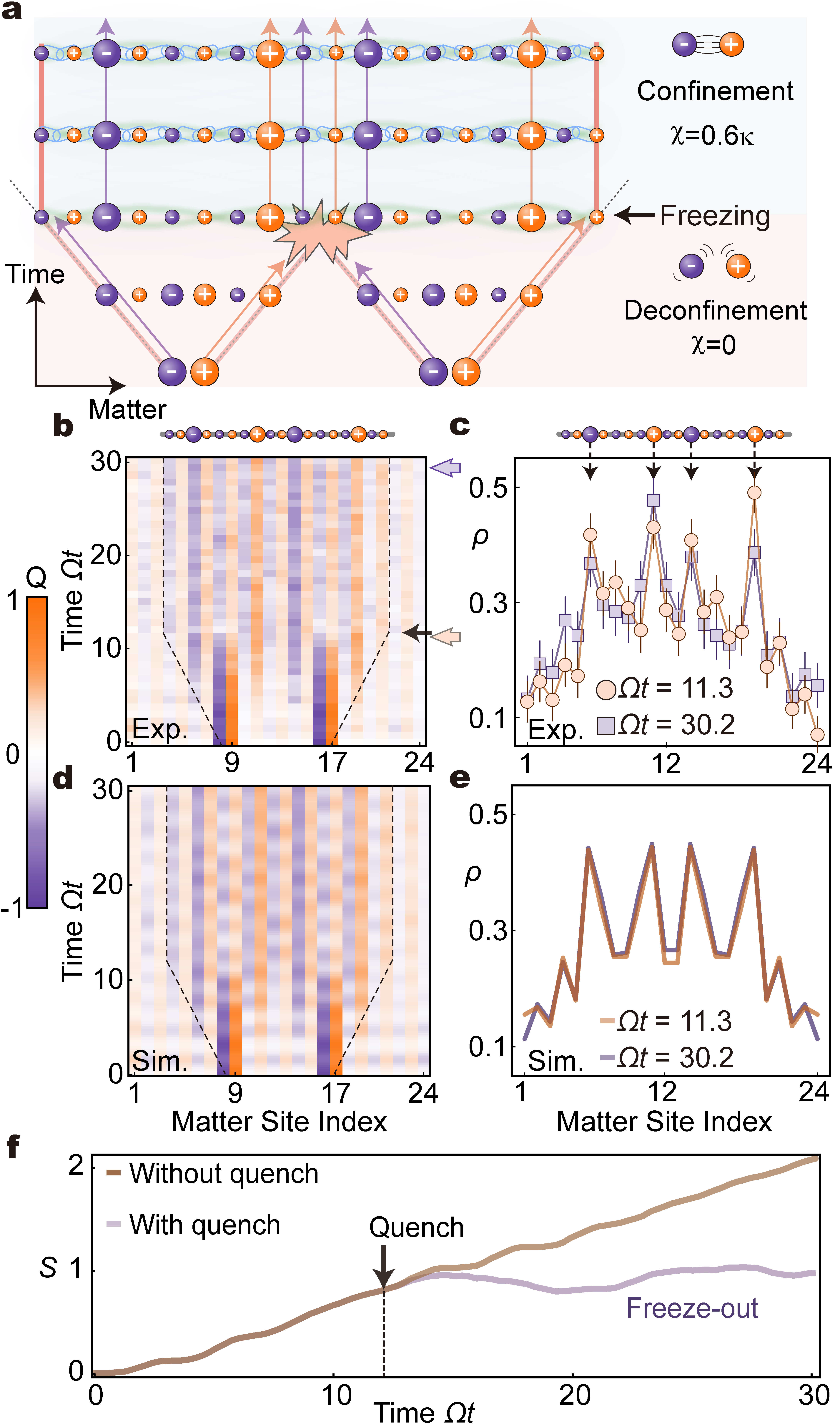}
  \caption{\textbf{Quantum freeze-out dynamics.} 
  \textbf{a}, Illustration of the quantum freeze-out. The system is initially prepared in the deconfinement phase, allowing free propagation of charges. At the onset of collision ($\Omega t=12.1$, indicated by the black arrow), the string tension is abruptly increased from $\chi=0$ to $0.6\kappa$, which freezes the large entangled scattering state and halts further evolutions. \textbf{b, c}, Experimentally measured site-resolved evolution of dynamical charge $Q_i(t)$ (\textbf{b}) and particle density $\rho_i(t)$ (\textbf{c}). The parallel propagation fronts (dashed lines, \textbf{b}) and the preserved transient collision configuration (\textbf{c}) are clearly observed following quantum quench. Inset: Charge distribution with magnitude-scaled circles. \textbf{d, e}, Numerical simulations corresponding to the experimental observations in \textbf{b, c}, respectively. Excellent agreement between experimental data and numerical predictions is found. 
  \textbf{f}, Numerical simulations of the half-chain entanglement entropy $S(t)$ with (purple line) and without (brown line) quench using experimental parameters (Methods). The rapid entropy growth is suppressed when the string tension is quenched after $\Omega t=12.1$, highlighting the freezing of the scattering state. 
  }
  \label{Fig:5_Freeze-frame dynamics}
\end{figure}

The dynamics in the strong confinement regime ($\chi = 0.6\kappa$) are probed by tuning the topological $\theta$ angle via staggered detuning $\delta$, breaking the vacuum degeneracy and stabilizing the charges.
The charges exhibit spatial localization throughout evolution, quantified by site-resolved dynamical charge $Q_i(t)$ (Fig.~\ref{Fig:4_Meson_dynamics}\textbf{a}) and density $\rho_i(t)$ (Fig.~\ref{Fig:4_Meson_dynamics}\textbf{c}), with minimal background fluctuations. 
Reducing $\chi$ while maintaining confinement yields moderate bounded wave packet spreading, as discussed in-depth in Methods.

Transitioning to deconfinement ($\chi = 0$) fundamentally alters the dynamics: The linear confining potential is replaced by the vanishing of the string tension, manifesting as ballistic propagation of decoupled constituent charges~\cite{surace_lattice_2020, Karpov_Spatiotemporal_2022}—positive (negative) charges moving rightward (leftward)—along light-cone trajectories (Fig.~\ref{Fig:4_Meson_dynamics}\textbf{b} dashed lines), exhibiting pronounced spatial broadening with no energy cost for charge separation.
Consequently, as right-moving positive charge (from the left excitation) and left-moving negative charge (from the right excitation) propagate outward, their wave packets overlap at $\Omega t \approx 11$. This collision dynamics generates a diamond-shaped interference pattern, characteristic of (1+1)D quasi-elastic scattering~\cite{surace_lattice_2020}. The transient central peak at $\Omega t = 19.6$ evolves into symmetric double peaks at $\Omega t = 29.4$ (Fig.~\ref{Fig:4_Meson_dynamics}d), as scattered particles continue to propagate beyond the overlapping region, in excellent agreement with numerical simulations (Fig.~\ref{Fig:4_Meson_dynamics}\textbf{e}-\textbf{f}). Concurrently, density peaks exceeding vacuum background fluctuations appear at the left (right) edges, corresponding to uncollided negative (positive) charges propagating freely outward.

The spatio-temporal control of Hamiltonian parameters allows us to explore a combined dynamics, where the value of the $\theta$ angle is changed abruptly along scattering---and thus, interactions between mesons change.  Specifically, we probe early scattering stage dynamics by initializing the system in the deconfinement regime ($\chi=0$, $m = 1.5\kappa$), where free charge motion enables wave packet overlap. At collision onset ($\Omega t^*=12.1$), we abruptly quench to strong confinement ($\chi=0.6\kappa$, $m = 1.8\kappa$) (Fig.~\ref{Fig:5_Freeze-frame dynamics}\textbf{a}).
This double quench protocol is a highly non-equilibrium, non perturbative process: as such, a naive expectation would be that of a continuous evolution towards a thermal state at long times. Instead, we observe that the transient scattering configuration is approximately conserved: this is evidenced by near-identical density profiles at $\Omega t=11.3$ (pre-quench) and $30.2$ (post-quench) with high fidelity, alongside stabilized charge distributions (insets, Fig.~\ref{Fig:5_Freeze-frame dynamics}\textbf{b}-\textbf{e}). We denote the result of this double quench as dynamical freeze-out.

We note that such dynamical freezing is not trivial at all, given the fact that the scattering state is very entangled~\cite{pichler_real-time_2016}, yet the growth of half-chain entanglement entropy is strongly suppressed after the string tension quench (Fig.~\ref{Fig:5_Freeze-frame dynamics}\textbf{f}),  rendering simple energetic reasoning not necessarily applicable. 
Extending this methodology, particle scattering and the associated nonequilibrium dynamics induced by deconfinement-to-confinement quenches at different fermion masses are also systematically explored (Methods).
Dynamical freeze-out is an emergent phenomenon, which we expect might have broader connection to other dynamical aspects---even beyond the dichotomy of deconfining-confining dynamics. Moreover, it is reminiscent of freeze-out techniques in the context of heavy-ion dynamics, where resulting scattering states admit descriptions in terms of equilibrium.
The flexibility of halting dynamics and switching interactions on and off in Rydberg experiments provides thus an ideal testbed to access the validity of freeze-out comparison between experiment and equilibrium theoretical modelling, similar to what is done in the context of heavy-ion collisions and lattice QCD~\cite{borsanyi2013freeze, Heinz:2007in, PhysRevLett.109.192302}.

\vspace{.5cm}

\noindent\textbf{\large Discussion and outlook}

\noindent In summary,  we have observed real-time scattering and freeze-out dynamics in a (1+1)D U(1) lattice gauge theory with a programmable Rydberg quantum simulator. In particular, we have demonstrated quench-driven confinement-deconfinement transitions characterized by string fragmentation and symmetry restoration. Furthermore, we tracked real-time charge scattering processes, revealing characteristic diamond-shaped interference patterns, and captured transient scattering states via a quantum freeze-out protocol that quenches to confinement at collision onset. Our approach enables future exploration of fundamental non-equilibrium phenomena in gauge theories—including
dynamical quantum phase transitions~\cite{Huang_Dynamical_2019, Zache_Dynamical_2019}, weak ergodicity breaking~\cite{Desaules_Weak_2023, desaules_ergodicity_2024} and emergent hydrodynamics~\cite{Berges_QCD_2021}—through high-precision Rydberg Hamiltonian control. The demonstrated quantum freeze-out dynamics provides a template for probing ultra-fast dynamics in complex scattering processes beyond perturbative regimes. Combined with scalable wave-packet engineering, this positions Rydberg quantum simulator as a promising platform for accessing high-energy dynamics that are challenging for classical computation~\cite{bauer_quantum_2023}.

In addition, realization of multi-body Wilson plaquette operators~\cite{glaetzle2014quantum,dai_four-body_2017, meth_simulating_2025} will extend Rydberg quantum simulators of LGT to higher dimensions, enabling explorations of emergent phenomena like monopole condensation—a conjectured mechanism underpinning quark confinement in QCD. Concurrently, simulating non-Abelian gauge theories (such as those based on SU(2) and SU(3)) demands hybrid architectures merging analog Hamiltonian engineering with digital quantum circuits to implement non-commuting group operations~\cite{gonzalez-cuadra_hardware_2022, zache_fermion-qudit_2023}. Progress along these fronts shall enable first-principles studies of non-Abelian gauge theories with high temporal resolution and tunability, including string fragmentation and hadronization in real-time multistage hadronic collisions and cosmological settings~\cite{gross_50_2023}.

\vspace{.6cm}
\noindent \textit{Note}.---After the completion of our experiments and during the preparation of this manuscript,  we became aware of two related and complementary studies that appeared on arXiv, where observations of  hadron scattering on superconducting (ibm\_marrakesh)  \cite{Schuhmacher2025Observation} and trapped-ion (IonQ Forte)  \cite{Davoudi2025Quantum} quantum digital simulators are reported, respectively. 

\vspace{.5cm}
\noindent\textbf{\large{}Acknowledgements} 
The authors acknowledge insightful discussions with Bing Yang, Tao Shi, Da-Wei Wang, Weibin Li, and Weikang Li.
The numerical simulations in this work are completed in the HPC Platform of Huazhong University of Science and Technology. 
This work was supported by the National Key Research and Development Program of China (Grant No.~2021YFA1402003), the National Science and Technology Major Project of the Ministry of Science and Technology of China (Grant No.~2023ZD0300901), the National Natural Science Foundation of China (Grant Nos.~T2225008, 12075128, 12374329, U21A6006, and 123B2072), the Science and Technology Commission of Shanghai Municipality, China (Grant No. 25LZ2601000), the Innovation Program for Quantum Science and Technology (Grant No.~2021ZD0302203),  and the Tsinghua University Dushi Program. M.\,D. was partly supported by the QUANTERA DYNAMITE PCI2022-132919, by the EU-Flagship programme Pasquans2, and the ERC Consolidator grant WaveNets. C.L. and D.-L.D. acknowledge in addition support from the Shanghai Qi Zhi Institute Innovation Program SQZ202318.

\vspace{.3cm}
\noindent\textbf{Author contributions} D.-S.X., P.Z., and H.-X.L. carried out the experiments under the supervision of L.L.. D.-S.X. performed the numerical simulations with the support from C. L. and P.Z.. C. L., P.Z., D.Y., M.D., and D.-L.D. conducted the theoretical analysis. D.-S.X., P.Z., H.-X.L., Y.-W.Z., K.Z., and B.X. contributed to the experimental setup. P.Z., C. L., D.-S.X., D.Y., M.D., D.-L.D., and L.L. wrote the manuscript with input from all authors.

\vspace{.3cm}
\noindent\textbf{Competing interests}  All authors declare no competing interests.

\bibliographystyle{naturemag}
\bibliography{library.bib}

\vspace{.5cm}
\noindent\textbf{\large{Methods}}
\setcounter{figure}{0}
\renewcommand{\theHfigure}{A.Abb.\arabic{figure}}
\renewcommand{\figurename}{Extended Data Fig.}

\noindent\textbf{\large{}Experimental details}

\noindent\textbf{Programmable atomic array and Rydberg excitation.} 
Our experiment utilizes a programmable Rydberg atom array that operates as an analogue quantum simulator~\cite{fang_Peobing_2024, shaw_benchmarking_2024, Chen_Interaction_2025, manovitz_quantum_2025, Chen_Spectroscopy_2025}. 
Individual $^{87}$Rb atoms are captured from a laser-cooled atomic ensemble into a tightly focused two-dimensional array of optical tweezers, generated by diffracting an 808-nm laser beam from a spatial light modulator (SLM). A pair of orthogonally oriented acousto-optic deflectors (AODs) is used to rearrange the atoms into a defect-free one-dimensional chain with uniform spacing $a$, where the nearest-neighbor van der Waals interaction between Rydberg atoms reaches a strength of $V_{i,i+1}\approx 2\pi\times7.2 \text{MHz}$. After rearrangement, atoms are further cooled in the tweezers to a temperature of approximately $10~\mu\text{K}$ and then optically pumped into the ground state $\lvert g\rangle = \lvert5S_{1/2}, F=2, m_F=2\rangle$.

Coherent coupling between the ground state $\lvert g \rangle$ and a highly excited Rydberg state $\lvert r \rangle$ is implemented via a two-photon transition driven by counter-propagating 780-nm and 480-nm laser beams, with single-photon Rabi frequencies $\Omega_{780}$ and $\Omega_{480}$. Both beams are detuned from the intermediate state $\lvert5P_{3/2}\rangle$ by $\Delta_I = 2\pi \times 1.2~\text{GHz}$, resulting in an effective two-photon Rabi frequency $\Omega = \Omega_{780} \Omega_{480} / (2\Delta_I)$.

The 480-nm Rydberg excitation beam is generated via second-harmonic generation (SHG) from a 960-nm seed laser. Both the 780-nm and 960-nm lasers are frequency-stabilized using the Pound–Drever–Hall (PDH) technique, referencing a high-finesse ultra-low-expansion (ULE) optical cavity. An additional slow feedback loop, applied outside the main PDH circuit, is used to compensate for long-term frequency drifts of both lasers.
To initiate the experimental sequence with precise timing, an electro-optic modulator (EOM) is employed to gate the 780-nm excitation beam with a timing resolution of approximately 10~ns ($\Omega t < 0.1$). To suppress shot-to-shot fluctuations in the effective Rabi frequency, we implement a sample-and-hold scheme using acousto-optic modulators (AOMs) on both Rydberg excitation legs.
The resulting setup supports long coherence times ($T_2^* \sim 11$ $\mu s$ and $\Omega T_2^* \sim 80$) for Rydberg Hamiltonian evolution, providing the foundation for high-fidelity quantum control and state preparation~\cite{xiang_observation_2024}.

\noindent\textbf{Preparation of Meson-like excitation.} Following initialization, we prepare initial meson-like states for the study of scattering dynamics. Under the lattice gauge theory mapping, a meson corresponds to a pair of localized domain defects embedded in a $\mathbb{Z}_2$-ordered background. A representative meson-like state takes the form (using a 13-atom chain as an example) $\lvert \text{rgrgr\textbf{ggg}rgrgr} \rangle$ while the state with two meson-like excitations is represented by $\lvert \text{r\textbf{ggg}rgrgr\textbf{ggg}r}\rangle$. Due to the inherent spatial asymmetry and local structure of these excitations, high-fidelity preparation using only amplitude- or phase-modulated global Rydberg pulses remains challenging. Nonetheless, high-fidelity many-body state preparation is essential for probing the non-equilibrium dynamics.

\begin{figure*}[h]
  \includegraphics[width=\textwidth]{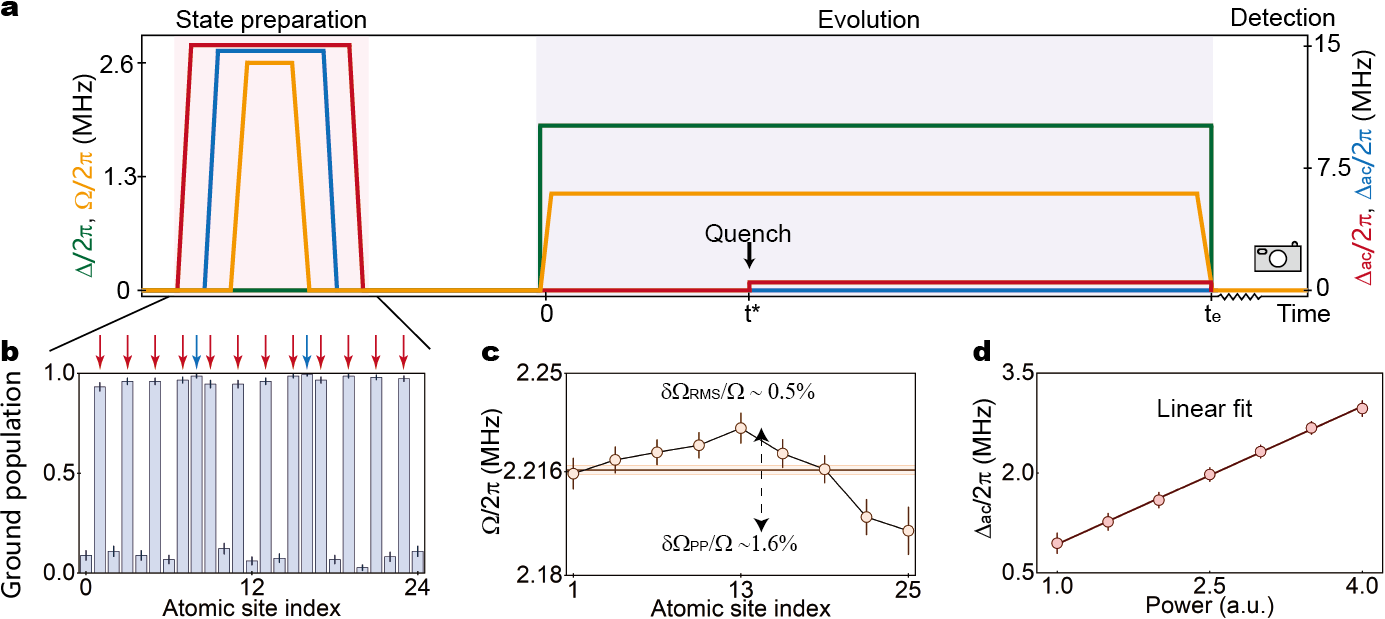}
  \caption{\textbf{Experimental setup and timing sequence.} 
  \textbf{a}, Timing sequence of the string tension quench experiment. The sequence begins with preparation of a meson-like state by combining a high-fidelity global Rydberg $\pi$ pulse (orange line, Rabi frequency $\Omega/2\pi \approx 2.6~\mathrm{MHz}$) with site-dependent addressing beams (red and blue lines) . The addressing beams produce large differential AC Stark shifts that detune selected atoms from the Rydberg transition, as illustrated in the inset of \textbf{b}. Red (blue) arrows indicate the addressing configuration used for initializing the $\mathbb{Z}_2$-ordered vacuum (creating meson-like excitations). The system subsequently evolves under the lattice gauge theory Hamiltonian, with the positive mass set by a global blue detuning (green line). At the quench time $t^*$, the addressing beam used for preparing the $\mathbb{Z}_2$ vacuum (red line) is rapidly activated to quench the string tension. After the evolution period ($t > t_e$), atomic states are detected via site-resolved fluorescence imaging.
 \textbf{b}, Measured atomic ground-state populations after state preparation (raw experimental data), showing clear site-resolved meson-like structure. A preparation fidelity of 51(5)\% is obtained after correcting for detection errors.
 \textbf{c}, Measured inhomogeneity of the Rabi frequency across the array. The peak-to-peak variation ($\delta \Omega_\mathrm{PP} / \Omega$)is approximately 1.6\%, with a root-mean-square (RMS) fluctuation ($\delta \Omega_\mathrm{RMS} / \Omega$) of 0.5\%.
 \textbf{d}, Measured AC Stark shift as a function of the radio-frequency (RF) power applied to the acousto-optic modulator (AOM) that diffracts the addressing beam. A clear linear dependence is observed, which is used for initial calibration of the light shifts required during confinement evolution.
}
  \label{EDFig:experimental_setup}
\end{figure*}

To this end, we combine global Rydberg excitation with spatially selective light shifts to prepare the meson-like initial states. Two far-detuned 795-nm laser beams (detuned by approximately 15~GHz from the D1 line) are shaped using a single spatial light modulator (SLM) to generate programmable light-shift patterns. One beam is used to prepare a staggered light pattern for preparing the $\mathbb{Z}_2$-ordered vacuum background, while the other selectively addresses additional sites to introduce domain walls corresponding to the meson-like excitations (inset of Extended Data Fig.~\ref{EDFig:experimental_setup}\textbf{b}). All addressing beams induce local AC-Stark shifts of approximately $2\pi \times 15~\text{MHz}$ on selected atoms, while maintaining a low photon scattering rate of $2\pi \times 5~\text{kHz}$. During a subsequent global $\pi$-pulse on the two-photon Rydberg transition, the addressed atoms are shifted out of resonance and remain in the ground state, whereas unaddressed atoms are resonantly transferred to the Rydberg state. This procedure results in the preparation of single- or multiple meson-like excitations depending on the spatial pattern and enables flexible and scalable preparation of highly structured many-body initial states. In a 25-site chain, we achieve a meson-like state fidelity of 51(5)\% (detection errors corrected), with a near-linear degradation of fidelity observed as the system size increases.

\noindent\textbf{Single-site addressing and evolution.} Exploration of the (1+1)D lattice gauge theory (LGT) phase diagram becomes particularly interesting when tuning the topological $\theta$ angle, which alters the vacuum structure of the background electric field, leading to confinement–deconfinement transitions and the emergence of string-breaking dynamics. In the effective Hamiltonian, the $\theta$ term is implemented via a staggered local detuning of the form $\delta_i = (-1)^i \delta$, where $i$ is the atomic site index.

Experimentally, this alternating detuning pattern is realized using one of the same 795-nm addressing beams employed during meson-state preparation. To enable both $\mathbb{Z}_2$-ordered vacuum initialization and Hamiltonian evolution, we dynamically switch the light shifts using an AOM within a single experimental sequence (Extended Data Fig.~\ref{EDFig:experimental_setup}\textbf{a}). Each addressed atom experiences a local AC-Stark shift $\Delta_\mathrm{ac}$ on the ground state, and further applying a global detuning of $\Delta_\mathrm{ac}/2$ yields the desired staggered detuning profile $\delta_i = (-1)^i \Delta_\mathrm{ac}/2$.

Tuning of the $\theta$ term and coherent evolution under local addressing require careful suppression of decoherence and optimization of light-shift uniformity. Dephasing primarily originates from intensity fluctuations in the addressing laser beams and atomic position uncertainties. To mitigate these effects, we employ addressing beams with a Gaussian waist of $3~\mu\mathrm{m}$—significantly larger than the atomic position uncertainty (about $300~\mathrm{nm}$)—to reduce sensitivity to atomic motion, and implement precision feedback loops to stabilize laser intensity and frequency. Additionally, spatial inhomogeneities in the light-shift pattern, caused by beam alignment drifts or optical crosstalk, can introduce variations in local detuning amplitudes $\delta$. To suppress such imperfections, we execute a two-step calibration protocol: initial alignment is achieved through achromatic imaging of atomic fluorescence and addressing beam profiles on an electron multiplying charge-coupled device (EMCCD) camera, followed by iterative SLM hologram adjustments based on measured local light shifts to optimize spatial uniformity and crosstalk suppression. 

\noindent\textbf{Experimental parameters} Under the experimental configuration described above, the effective Hamiltonian governing the system reads:
\begin{align}
\hat{H}_{\mathrm{exp}} = & \frac{\Omega}{2} \sum_i \hat{\sigma}^x_i - \Delta_0 \sum_i \hat{n}_i - \nonumber\\& \Delta_{\mathrm{ac}} \sum_{i \in \text{odd}} \hat{n}_i+ \sum_{i<j} \frac{C_6}{(a |i-j|)^6} \hat{n}_i \hat{n}_j.    
\end{align}
It can be equivalently rewritten as:
\begin{align}
\hat{H}_{\mathrm{exp}} = &\frac{\Omega}{2} \sum_i \hat{\sigma}^x_i - (\Delta_0 + \delta) \sum_i \hat{n}_i + \nonumber \\ & \delta \sum_i (-1)^i \hat{n}_i + \sum_{i<j} \frac{C_6}{(a |i-j|)^6} \hat{n}_i \hat{n}_j.    
\end{align}
Here, $\delta = \Delta_{\mathrm{ac}}/2$ is the staggered detuning, $\Delta_0$ denotes the global detuning of the Rydberg laser. $C_6$ is the van der Waals interaction coefficient, and $a$ is the lattice spacing. This Hamiltonian matches the form of the Rydberg Hamiltonian $H_{\mathrm{R}}$ introduced in the main text, with $\Delta = \Delta_0 + \delta$ and $V_{ij} = C_6 / (a |i - j|)^6$.

Throughout all measurements presented in the main text, the atomic array is configured such that the Rydberg blockade radius $R_b$ exceeds the lattice spacing $a$, with $R_b / a \gtrsim 1.3$. This condition ensures that Rydberg interactions effectively suppress the simultaneous excitation of nearest-neighbor atoms, thereby enforcing the Gauss law constraint.

For the measurements shown in Figs.~\ref{Fig:2_confinement_and_deconfinement_phases}, \ref{Fig:4_Meson_dynamics} and \ref{Fig:5_Freeze-frame dynamics} of the main text, we employ a Rabi frequency of $\Omega = 2\pi \times 1.2~\mathrm{MHz}$, with the chosen Rydberg state $\lvert 68D_{5/2}, m_j = 5/2 \rangle$ and a lattice spacing of $a = 7.2~\mu\mathrm{m}$, yielding $R_b / a = 1.35$. For the data presented in Fig.~\ref{Fig3:Quench_dynamics} of the main text, we use a Rabi frequency of $\Omega = 2\pi \times 1.5~\mathrm{MHz}$ and the Rydberg state $\lvert 72D_{5/2}, m_j = 5/2 \rangle$, with $a = 7.0~\mu\mathrm{m}$, resulting in $R_b / a = 1.3$. Across all configurations, the time step $\Delta T$ for stroboscopic measurements is chosen such that $\Omega \Delta T = 0.75$.

The long-range tail of the Rydberg interaction can be treated as an effective energy offset, which modifies the control of the mass parameter in the mapped lattice gauge theory. Experimentally, we compensate for this offset by introducing a small global laser detuning $\Delta \sim V_{i, i+1} / 32$ to establish the effective resonance point at $\Delta_0 = 0$. The mass parameter is then controlled by incrementally increasing the blue detuning of the Rydberg excitation laser, achieving $\Delta_0 > 0$. This approach enables the system to approximate the PXP model while maintaining precise control over the effective mass.

To probe confinement dynamics, we modulate local staggered detuning $\delta$ by tuning the radio-frequency (RF) power delivered to the AOM for the addressing beams. The relationship between $\delta$ and RF power is experimentally calibrated and exhibits predominantly linear dependence (Extended Data Fig.~\ref{EDFig:experimental_setup}d). This calibration provides an initial reference for setting $\delta$, followed by fine adjustment of the local AC-Stark shift ensuring precise control of the detuning.

To explore the dynamical confinement–deconfinement phase transition, we perform quenches of the staggered detuning. In the confinement-to-deconfinement transition shown in Fig.~\ref{Fig3:Quench_dynamics} of the main text, the local AC-Stark shift is initially set to $1.8\Omega = 2\pi \times 2.7~\mathrm{MHz}$, corresponding to $\delta = 0.9\Omega$. At time $t = 1.2~\mu\mathrm{s}$ ($\Omega t = 11.3$), the addressing beam is abruptly turned off, implementing a dual quench from $(\Delta_0, \Delta_{\mathrm{ac}}) = (0, 2\pi \times 2.7~\mathrm{MHz})$ to $(0, 0)$. To achieve fast switching, the addressing laser is focused into the AOM with a beam waist of approximately 100 $\mu$m, yielding an optical response time of $\sim$20 ns ($\Omega t \sim 0.15$)—much shorter than the experimental time step.

In the reverse process, the deconfinement-to-confinement quench experiment (Fig.~\ref{Fig:5_Freeze-frame dynamics} of the main text), the addressing beams are activated at $t = 1.6~\mu\mathrm{s}$, introducing a local AC-Stark shift of $0.6\Omega = 0.72~\mathrm{MHz}$ and quenching the system into the confined phase. Consequently, the effective detuning changes from $\Delta_0 + \delta = 1.5\Omega$ to $1.8\Omega$, which increases the mapped mass parameter. This increased mass suppresses vacuum fluctuations and improves the visibility of confinement dynamics.

\noindent\textbf{Experimental measurements.}
At the end of each experimental Hamiltonian evolution, site-resolved atomic fluorescence is collected onto an EMCCD camera, enabling atomic state readout via fluorescence imaging (Extended Data Fig.~\ref{EDFig:experimental_setup}\textbf{a}). From these measurements, $\langle \sigma_i^z(t) \rangle$ values are directly measured. The electric field $\langle E_i(t) \rangle = (-1)^i \langle \sigma_i^z(t) \rangle$ is obtained, where $i$ denotes the atomic (gauge) site index. Correspondingly, the particle number density at matter site $j$ is $\langle \rho_j(t) \rangle = (-1)^{j-1} \big( \langle E_j(t) \rangle - \langle E_{j-1}(t) \rangle \big)$, while the dynamical charge is $\langle Q_j(t) \rangle = \langle E_j(t) \rangle - \langle E_{j-1}(t) \rangle$, establishing $\langle \rho_j(t) \rangle = \left| \langle Q_j(t) \rangle \right|$ as presented in the main text. These transformations exhibit well-preserved compliance with Gauss's law, guaranteed by strong nearest-neighbor Rydberg blockade. All data in the main text represent raw measurements without post-processing.

In Fig.~\ref{Fig:2_confinement_and_deconfinement_phases}\textbf{c--e} of the main text, the mass dependence of confinement strength is investigated by comparing steady-state bulk electric flux values extracted from experimental data at different masses. To determine these steady-state values, the $\langle E_\mathrm{B}(t) \rangle$ time traces were fitted with a damped oscillation function: $\langle E_\mathrm{B}(t) \rangle = E_\mathrm{s} + E_\mathrm{o} e^{-\gamma t} \cos(\omega t + \phi_0)$, where $E_\mathrm{s}$ represents the steady-state value as $t \to \infty$. Fits converge well to the data (Fig.~\ref{Fig:2_confinement_and_deconfinement_phases}\textbf{c--e}).

\noindent\textbf{Experimental error analysis.}
Although the raw experimental data presented in the main text robustly capture all key physical phenomena and dynamical structures of interest in (1+1)D lattice gauge theory, we nevertheless provide an analysis of relevant experimental imperfections. These imperfections do not compromise the conclusions of this work but serve as a technical reference for future optimization efforts.

A primary source of error arises from the preparation of the initial meson-like states. Prior to dynamical evolution, the atom array is configured into a many-body state encoding a meson-like excitation. Although the fidelity of this state preparation reaches a high level, errors—most commonly single-spin-flip defects—can occur. These errors introduce stray pairs of positive and negative charges at unintended locations, which undergo slow thermalization. For scattering processes requiring a pristine vacuum background to resolve subtle signals, such initial-state imperfections could influence background fluctuations, though the primary scattering signatures remain robustly resolvable.

Decoherence constitutes an additional limiting factor for both confinement and deconfinement dynamics, arising from laser phase noise, finite atomic temperature, and technical imperfections. Specifically, phase noise from our excitation lasers exhibits a servo bump peak near $f_{\mathrm{servo}} \approx 300~\mathrm{kHz}$, attributed to the finite bandwidth of the feedback circuit. The broad spectral distribution of this servo bump substantially limits coherence for Rabi frequencies within its spectral range. Furthermore, the PDH locking of the seed laser (used for blue-light generation) introduces intensity noise predominantly in the servo bump frequency band. The low-frequency components of this intensity noise (below $\sim\Omega/2\pi$) contribute significantly to decoherence. Consequently, simply increasing the Rabi frequency does not always reduce decoherence; instead, we optimize $\Omega/2\pi \approx 5 f_{\mathrm{servo}}$ to balance phase and intensity noise effects. In confinement dynamics experiments, fluctuations in the local light shift also cause additional dephasing, manifesting as stochastic variations in the effective detuning of $\sim$10 kHz per atom.
Spatial inhomogeneities in the Rydberg excitation laser result in a Rabi frequency variation of $\delta \Omega_{PP} / \Omega \approx$1.6\% (peak-to-peak) and $\delta \Omega_{RMS} / \Omega \approx$0.5\% (root-mean-square) across the array (Extended Data Fig~\ref{EDFig:experimental_setup}\textbf{c}). Furthermore, atomic position uncertainties of approximately 0.3~$\mu m$ during evolution induce fluctuations in interaction strengths. These effects collectively lead to deviations from ideal many-body dynamics.

Finally, measurement imperfections arising during fluorescence imaging at the end of each experimental run also contribute to the experimental errors. For single atoms, we measure a detection error rate of approximately $1\%$  for the Rydberg state (attributed to finite state lifetime) and approximately ~2\% for the ground state (primarily caused by atom loss).

\vspace{.5cm}
\noindent\textbf{Mapping between quantum link model and Rydberg Hamiltonian}

\noindent In our programmable atomic array, each atom is modeled as a two-level system comprising a ground state $\lvert g\rangle$ and a Rydberg state $\lvert r\rangle$. The strong van der Waals interaction between Rydberg atoms gives rise to the Rydberg blockade effect, which prohibits simultaneous excitation of neighboring atoms. Following Refs.~\cite{surace_lattice_2020, cheng_emergent_2024}, this blockade constraint can be interpreted as an emergent $U(1)$ gauge invariance, analogous to Gauss's law in electrodynamics.
The system is governed by contributions of two terms: the coherent laser driving and the Rydberg interaction. The interaction term
\begin{equation}
\sum_{i<j} V_{ij} \hat{n}_i \hat{n}_j,\quad V_{ij} \sim C_6/(a|i - j|)^6    
\end{equation}
describes the van der Waals repulsion between atoms in the Rydberg state, where $\hat{n}_i = \lvert r_i\rangle\langle r_i\rvert$ is the number operator. The laser driving term 
\begin{equation}
\frac{\Omega}{2} \sum_i  \hat{\sigma}_x^i - \sum_i (\Delta + \delta_i)\hat{n}_i,   
\end{equation}
is characterized by the two-photon effective Rabi frequency $\Omega$, a global detuning $\Delta$, and a local site-dependent detuning $\delta_i$. 

In our experimental configuration, the nearest-neighbor interaction $V_{i,i+1}$ significantly exceeds all other energy scales: $\Omega$, $\Delta$, and $\delta_i$. The local detuning $\delta_i$ is particularly small and is therefore neglected in the following theoretical mapping. However, it plays an essential role in tuning the vacuum structure via the effective topological $\theta$ angle, as discussed below, without affecting the underlying gauge constraint. As a result of the dominant interactions, the doubly excited states $\lvert r_i r_j\rangle$ are energetically suppressed, and the dynamics are constrained to a subspace where neighboring Rydberg excitations are excluded. Projecting onto this constrained subspace yields the effective Hamiltonian

\begin{equation}
H_{\mathrm{PXP}} = \frac{\Omega}{2} \sum_i \hat{P}_{i-1} \hat{\sigma}_x^i \hat{P}_{i+1} - \Delta \sum_i \hat{n}_i,    
\end{equation}
where $\hat{P}_i = 1 - \hat{n}_i$ enforces the blockade constraint. This so-called PXP Hamiltonian~\cite{Lesanovsky_Interacting_2012} governs the dynamics of Rydberg arrays in the blockade regime.

\begin{figure*}[h]
  \includegraphics[width=\textwidth]{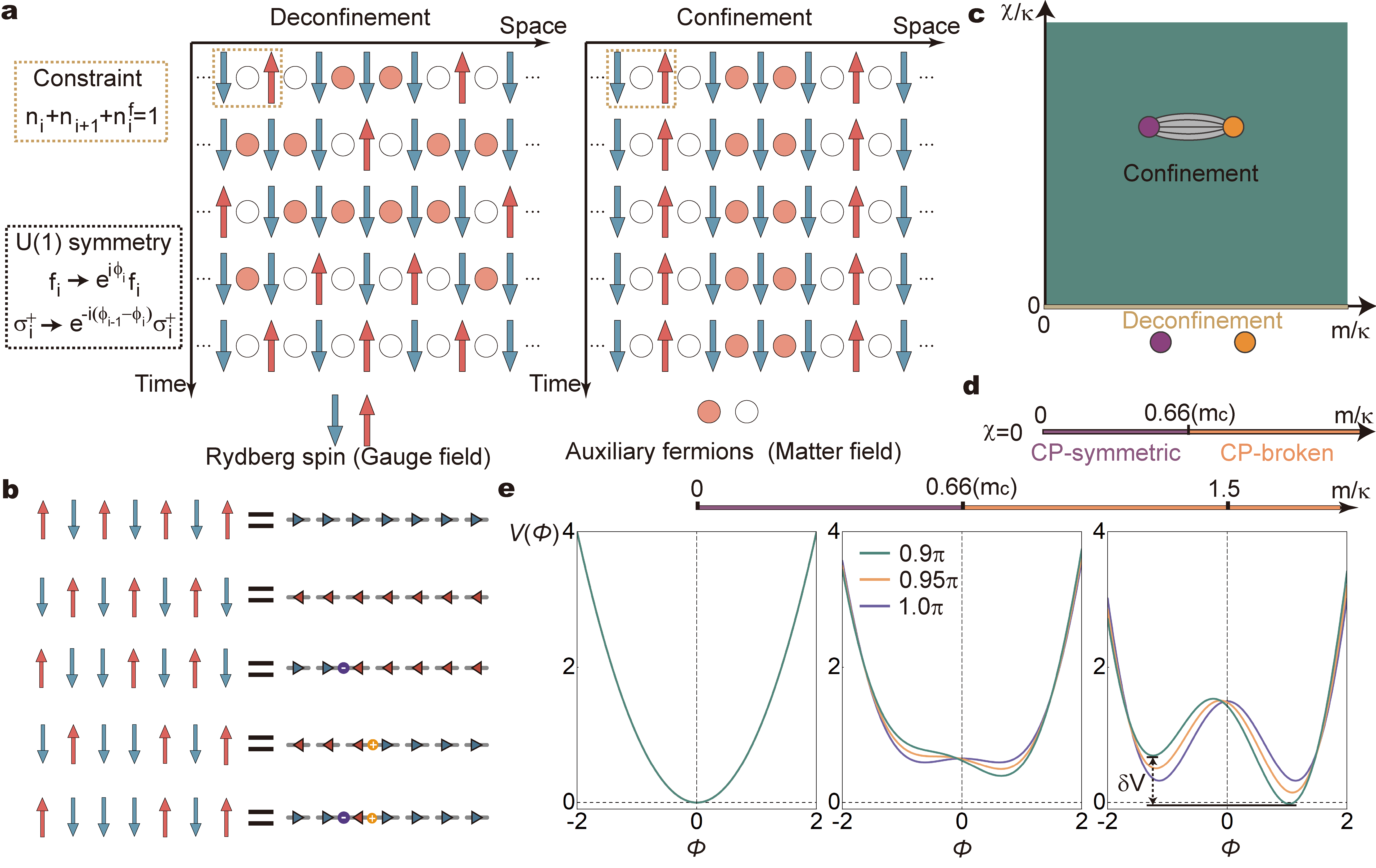}
  \caption{\textbf{Lattice gauge theory mapping to the Rydberg atom array.} 
  \textbf{a}, Realization of a $1+1$D U(1) lattice gauge theory (LGT) in a Rydberg atom array. In the PXP model regime, the Rydberg spins (arrows) represent gauge fields and auxiliary fermions (colored circles, matter fields) are introduced on the links between two spins. The Rydberg blockade between neighboring atoms enforces the Gauss law constraint $n_i + n_{i+1} + n_i^{\mathrm{f}} = 1$ and enables the U(1) gauge symmetry, thereby realizing an effective lattice gauge theory. The spacetime evolution of an initial meson-like state is illustrated for both the deconfinement and confinement phases.
  \textbf{b}, Representative patterns of the 1D atomic chain configurations. From top to bottom: positive electric field string (+E), negative electric field string (-E), negative charge excitation, positive charge excitation and negative-positive charge pair.
  \textbf{c--d}, Phase diagram of the $1+1$D U(1) LGT, depending on the fermion mass $m/\kappa$ and the effective string tension $\chi/\kappa$. In the confinement regime ($\chi \neq 0$, $\theta \neq \pi$), particles are bound by electric strings, whereas in the deconfinement regime ($\chi = 0$, $\theta = \pi$), particles move freely. A spontaneous charge-parity symmetry breaking emerges at $\chi = 0$ when $m > m_c$, with $m_c / \kappa = 0.66$ (\textbf{d}).
  \textbf{e}, Scalar potential $V(\phi)$ as a function of $m$ and $\theta$ (we set $c = 1$ and $e = 1$ in all calculations). In the deconfinement phase ($\theta = \pi$), a saddle point appears at $\phi = 0$, and for $m > m_c$ the potential develops two minima at $\pm \phi_0$, resulting in spontaneous charge-parity symmetry breaking. In the confinement phase ($\theta \neq \pi$), the minima become biased, resulting in a potential gap $\delta V$ that defines the cost of separating particle pairs. The gap increases as the mass becomes larger and as $\theta$ deviates from $\pi$, highlighting stronger confinement.
}
  \label{EDFig:Mapping}
\end{figure*}

The projection operators in the PXP Hamiltonian place a constraint on the full Hilbert space, that no two atoms within the Rydberg blockade radius can be simultaneously excited to the Rydberg state. This constraint can be reformulated by enlarging the original Hilbert space with the aid of the so-called auxiliary fermions: Between each pair of Rydberg blockaded atoms, we define a fictitious fermion that sits on the link between the two sites and is annihilated when either atom is in the Rydberg state (in other words, the ``Dirac sea'' of the fictitious fermions is fully occupied when all atoms are in the ground state). Since applying the annihilation operator twice kills the quantum state, a pair of atoms sharing the same fermionic link cannot both be in the Rydberg state, effectively enforcing the Rydberg blockade effect. Denoting the annihilation operator of the fermion on the link between site $i$ and $i+1$ by $f_i$, we can rewrite the PXP Hamiltonian as
\begin{equation}\label{eq:ham}
H = \sum_i \left[\frac{\Omega}{2}\left(\sigma_i^- f^\dagger_{i-1} f^\dagger_i + \sigma_i^+ f_{i-1} f_i\right) - \Delta n_i\right]
\end{equation}
where we have defined $\sigma^\pm_i = \sigma^x_i \pm {\rm i} \sigma^y_i$. In this enlarged Hilbert space with the auxiliary fermions, the original constraint $n_i+n_{i+1}\le 1$ becomes an equality (Extended Data Fig.~\ref{EDFig:Mapping}\textbf{a})
\begin{equation}\label{eq:conserve}
n_i + n_{i+1} + n^f_i = 1\,.
\end{equation}
Here $n^f_i$ is the number operator of the auxiliary fermion $n^f_i = f^\dagger_i f_i$. This equality constraint is a conservation law on the enlarged Hilbert space. It therefore implies the existence of a gauge symmetry. Indeed, the Hamiltonian as well as the constraint is invariant under the following gauge transformation
\begin{equation}
f_i \to e^{{\rm i}\phi_i} f_i,\qquad \sigma^+_i \to e^{-{\rm i}(\phi_{i-1}+\phi_i)} \sigma^+_i,    
\end{equation}
where we have defined a lattice field $\phi_i$ that parametrizes the gauge transformation. Since $|e^{{\rm i}\phi_i}|=1$, we have therefore arrived at a $U(1)$ gauge theory defined on a lattice with one spatial dimension (Extended Data Fig.~\ref{EDFig:Mapping}\textbf{a}). The conservation law Eq.~\eqref{eq:conserve} is the analogue of the Gauss law in the usual electrodynamics in three spatial dimensions. One can therefore identify the fermionic field $f_i$ as the matter field, and the spin field $\sigma_i^z$ as the gauge field, the analogue of the electric field $E_i$ in our one-dimensional theory. In particular, $\sigma^+_i$ plays the role of the parallel transporter $U_i$ that links the fermionic field at site $i$ to site $i+1$, and satisfies $[E_i,U_i]=U_i$. Using the constraint Eq.~\eqref{eq:conserve}, the Hamiltonian Eq.~\eqref{eq:ham} can be rewritten as
\begin{equation}\label{eq:ham2}
H = \sum_i \left[\frac{\Omega}{2}\left(f^\dagger_{i-1} U_i^\dagger f^\dagger_i + f_{i-1} U_i f_i\right) - (-1)^i \Delta n^f_i\right],
\end{equation}
where we have ignored an overall constant and the boundary terms coming from replacing $n_i$ with $n^f_i$. We observe that the generator of the $U(1)$ gauge transformation
\begin{equation}
G_i = E_{i-1} - E_i - n^f_i, 
\end{equation}
commutes with the Hamiltonian, $[G_i,H]=0$. This is the analogue of the Gauss law in our theory.

The Hamiltonian in Eq.~\eqref{eq:ham2} takes the form of a (1+1)D U(1) quantum link model, where the fermionic fields $f_i$ are interpreted as staggered fermions, which alternate between matter and anti-matter on even and odd lattice sites. A background electric field $E_i^{\mathrm{bg}}$ can be incorporated by shifting $E_i \to E_i + E_i^{\mathrm{bg}}$, which effectively introduces a topological $\theta$-term into the Hamiltonian. This term corresponds to a local staggered detuning $\delta_i = (-1)^i \delta$, which experimentally manifests as an additional contribution proportional to the local AC-Stark shift. By tuning $\delta$, we dynamically control the topological $\theta$ angle via the relation $\delta = J(\pi-\theta)/2\pi$~\cite{banerjee_atomic_2012,surace_lattice_2020,cheng_tunable_2022}, thereby enabling access to different vacuum structures and real-time confinement dynamics, as explored in the main text.

\vspace{.5cm}
\noindent\textbf{Schwinger model}

\noindent In the continuous limit, the parallel transporter $U_i$ becomes the covariant derivative $D_\mu=\partial_\mu -i e A_\mu$ with $\mu=0,1$, where $A_\mu$ is the continuous version of the lattice gauge field $\phi_i$, and our quantum link model becomes the theory of quantum electrodynamics (QED) in one spatial dimension, otherwise known as the Schwinger model, with the Hamiltonian
\begin{equation}
H_{\rm QED} = \int {\rm d}x \left[\psi^\dagger (i\gamma^\mu D_\mu-m)\psi - \frac{1}{4}F_{\mu\nu}F^{\mu\nu}\right],   
\end{equation}
where we have identified the mass of the fermionic field $m=\Delta$. Here, $\psi$ is the two-spinor that corresponds to the fermionic field $f$ discussed above, $\gamma^\mu$ are the $\gamma$ matrices in two dimensions, and $F_{\mu\nu}$ is the field strength tensor defined by $F_{\mu\nu} = \partial_\mu A_\nu - \partial_\nu A_\mu$. Since $F_{\mu\nu}$ is fully anti-symmetric, in $(1+1)$D it has exactly one independent component, which we identify as the electric field $E_i$ as discussed in the main text. In $(1+1)$D, one can use the epsilon tensor $\epsilon^{\mu\nu}$ to replace the fermionic degree of freedom $\psi$ with a bosonic field $\phi$, through the following relation
\begin{equation}
\psi^\dagger \gamma^\mu \psi = \epsilon^{\mu\nu} \partial_\nu \phi.    
\end{equation}

Substituting this relation into the $(1+1)$D QED Lagrangian and integrating over the gauge field $A_\mu$, one arrives at a bosonic theory that is equivalent to the original fermionic theory, with the following Hamiltonian~\cite{surace_lattice_2020}:
\begin{equation}\label{eq:Hb}
H_{\rm B} = \frac{1}{2}\int {\rm d}x\left[\dot{\phi}^2 + (\partial_x\phi)^2+e^2\phi^2 - cm\cos(2\phi-\theta)\right]
\end{equation}
Here, $e$, $c$, and $\theta$ are all constants related to $\Omega$ and $\Delta$. In particular, the last term in the bosonizied Hamiltonian is known as the topological $\theta$ term.  This term modifies the vacuum structure of the theory, leading to a $\theta$-dependent mass gap in the spectrum and therefore potential confinement and deconfinement of the charged fermion. The value of $\theta$ depends on the Rydberg system parameters $\Delta$ and $\delta$ as discussed in main text.

\vspace{.5cm}
\noindent\textbf{Confinement and deconfinement phases}

\noindent The presence of the $\theta$-dependent potential in the bosonizied theory $H_{\rm B}$ gives rise to a rich vacuum structure at non-zero fermion mass, where the location and degeneracy of the minima of the potential $V(\phi) = e^2 \phi^2-cm\cos(2\phi-\theta)$ depend sensitively on the value of $\theta$. This leads to phenomena such as charge-parity symmetry breaking and confinement-deconfinement transition, both of which are observable in our Rydberg-atom simulation platform. The vacua of the theory correspond to the values of $\phi$ that minimize the potential $V(\phi)$. Assuming $cm>0$, for $\theta=0$ there is a unique minimum at $\phi=0$, and the vacuum therefore prefers configurations with zero net electric field. In this regime, the energy cost of separating static charges grows linearly with distance due to the formation of a stable electric flux tube. This corresponds to the confining phase, where dynamical string breaking is strongly suppressed.

As $\theta$ increases, the minimal $\phi$ shifts away from zero and the vacuum acquires a nonzero net electric field, gradually transitioning the system into a less confined regime. As $\theta$ approaches $\pi$, the energy difference between the global minimum and the nearest local minimum decreases, vanishing exactly at $\theta = \pi$ (Fig.~\ref{Fig1:experimental_set_up}\textbf{e} and Extended Data Fig.~\ref{EDFig:Mapping}\textbf{e}), where the potential exhibits two degenerate global minima located at $\phi = \pm \phi_0/2$. In this regime, the potential is again symmetric (just like in the case of $\phi=0$), and the vacuum exhibits a two-fold degeneracy, corresponding to opposite orientations of the electric field. This symmetry structure is intimately connected to charge-parity symmetry breaking (see the next section), and gives rise to the the deconfinement dynamics as a result of enhanced particle-antiparticle pair production due to vanishing string tension.

In the case where the fermion mass is zero ($m = 0$), the cosine term in the potential $V(\phi)$ disappears entirely, reducing the theory to that of a free scalar field. In the absence of a confining potential, there is no energy cost associated with separating particle-antiparticle pairs, and the system is therefore in a deconfined phase characterized by unimpeded particle propagation and the absence of string tension.

\vspace{.5cm}
\noindent\textbf{Tuning of string tension}

\noindent In higher dimensional, quenched gauge theories, the Wilson loop operator $W(C)$, constructed as the trace of a product of link variables along a closed loop $C$, serves as an order parameter for confinement. In the confined phase, where we expect a linearly growing energy between a static positive-negative charge pair, the expectation value of the Wilson loop operator follows an area law $\langle W(C) \rangle \propto e^{-\chi \cdot A(C)}$, where $A(C)$ is the area enclosed by the loop and $\chi$ is the string tension. In the deconfined phase, it instead follows a perimeter-law, $\langle W(C) \rangle \propto e^{-k \cdot P(C)}$ where $P(C)$ is the perimeter of the loop, indicating that charges are free to separate at arbitrary distance with a cost that is approximately distance independent. Importantly, these behaviors are typically characterizing loops on a space-time space, including one imaginary time and one spatial direction.

In our 1D case, the area law is associated to a linear scaling $\langle W\rangle \propto e^{-\chi L T}$ where $L$ and $T$ are the spatial length and temporal extent of the rectangular Wilson ``loop''. In other words, we expect in 1D systems, the potential energy between two confining static charges will scale linearly $U(L)\propto \chi L$. The perimeter in 1D simply becomes a constant, so for the deconfining phase we expect the energy to be constant $U(L)\propto{\rm const}$. It is important to note that this picture, for static charges, is essentially classical, as no transverse gauge degree of freedom is present. We will thus refer to the quantity $\chi$ as effective string tension.

In our highly tunable Rydberg simulator, $\chi$ can be controlled by adjusting the topological $\theta$ angle via a staggered detuning $\delta_i = (-1)^i \delta$, with the relation $\delta = J(\pi - \theta)/2\pi$, as illustrated in Extended Data Fig.~\ref{EDFig:Mapping}\textbf{e} - see also Ref.~\cite{banerjee_atomic_2012}, where the model was introduced, and Ref.~\cite{surace_lattice_2020} for its relation to Rydberg couplings.

At $\theta = \pi$, the scalar potential $V(\phi)$ exhibits qualitatively distinct behavior depending on the fermion mass $m$. When $m = 0$, the potential has a single minimum at $\phi = 0$, corresponding to a charge-parity-symmetric phase with vanishing electric field. As the fermion mass increases beyond a critical value $m_c$, the potential develops two degenerate minima at $\phi = \pm \phi_0$ (Fig.~\ref{Fig1:experimental_set_up}\textbf{e} and Extended Data Fig.~\ref{EDFig:Mapping}), resulting in spontaneous charge-parity symmetry breaking (Extended Data Fig.~\ref{EDFig:Mapping}\textbf{d}). Crucially, in both cases, the vacuum degeneracy ensures that reversing the local electric field—i.e., creating a particle–antiparticle pair—incurs no energy cost. The resulting potential between two charges remains constant, $U(L) \propto \mathrm{const}$, leading to a vanishing string tension and deconfinement dynamics, as observed in Fig.~\ref{Fig1:experimental_set_up}\textbf{e}. This is the same dynamics of the continuum version of (1+1)D QED.

However, when $\theta$ is detuned away from $\pi$, this degeneracy is lifted, and the two lowest-energy electric field configurations acquire a finite energy splitting $\delta V$ (Extended Data Fig.~\ref{EDFig:Mapping}e). In this case, separating a particle–antiparticle pair over a distance $L$ creates a domain wall--or ``string"--of reversed electric field of length $L$ between the charges. The corresponding energy cost scales linearly, $U(L) \propto \chi L$, giving rise to a finite effective string tension with $\delta V \propto \chi$. The potential gap $\delta V$ increases as the topological $\theta$ angle deviates further from $\pi$, which is realized experimentally by applying a stronger staggered detuning $\delta$ (Extended Data Fig.~\ref{EDFig:Mapping}\textbf{e}), thereby resulting in a larger effective string tension $\chi$ and hence stronger confinement. Based on these considerations, the string tension is defined as $\chi = 2\delta$ in our work, as discussed in the main text.

Moreover, when $\chi$ is held fixed, the gap $\delta V$ increases with larger fermion mass $m$, as shown in Extended Data Fig.~\ref{EDFig:Mapping}\textbf{e}, highlighting stronger confinement--consistent with the behavior observed in Fig.~\ref{Fig:2_confinement_and_deconfinement_phases}\textbf{a} (lower panel) of the main text.

\vspace{.5cm}
\noindent\textbf{Scattering dynamics with varying string tension}

\noindent In our experiment, by tuning the string tension, we explored scattering dynamics across different confinement regimes, as mentioned in the main text. The initial state consists of two meson-like excitations embedded in a vacuum background, characterized by the dynamical charge configuration $Q_i = \sum_{j=8,9,16,17} (-1)^{i+1} \delta_{i,j}$. The system then evolves under the lattice gauge theory Hamiltonian at different string tensions, realized by setting the topological theta term $\chi$ from $0$ to $0.6 \kappa$, while fixing the mass parameter at $m = 1.5 \kappa$.

\begin{figure*}[h]
  \includegraphics[width=\textwidth]{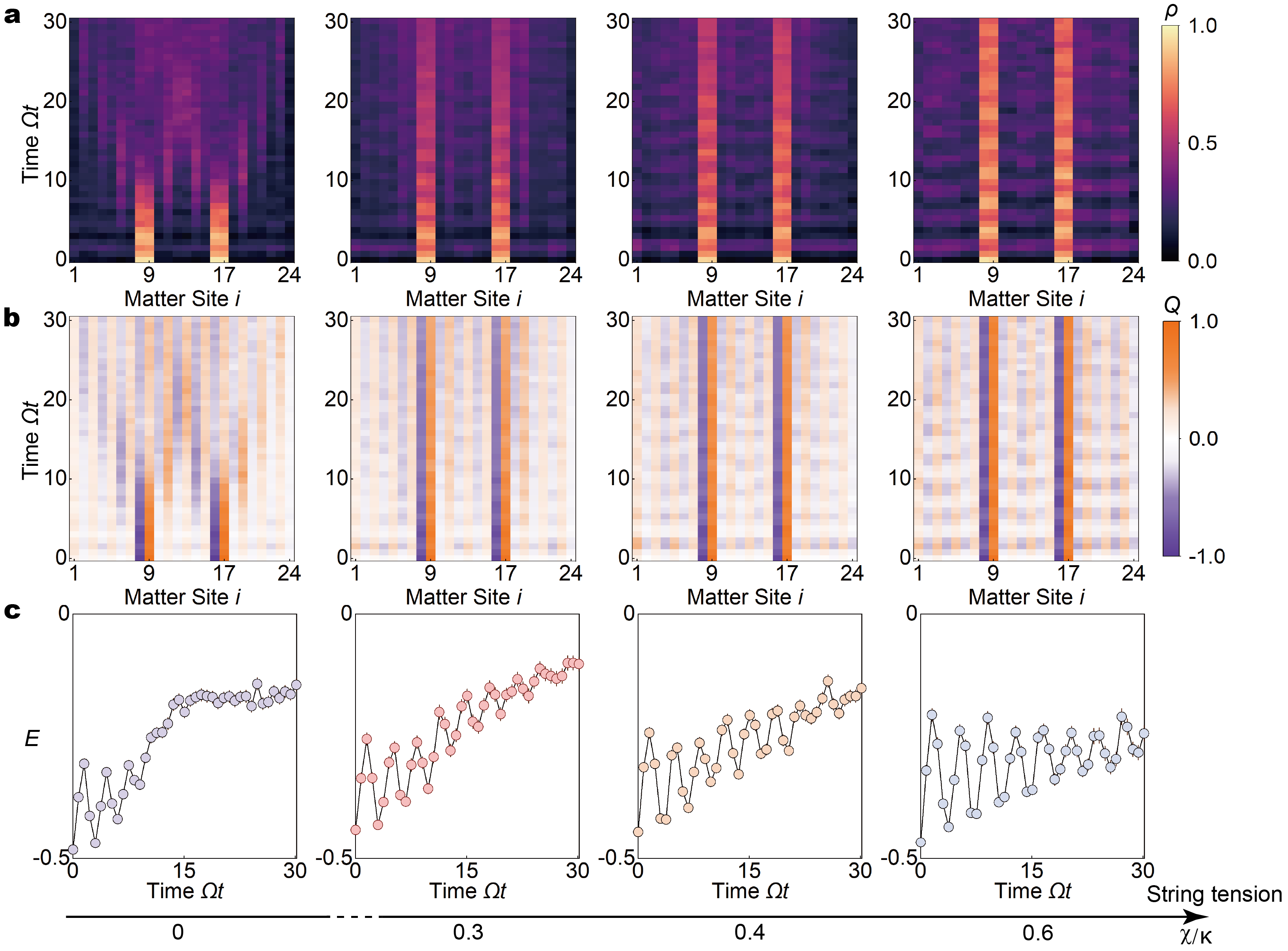}
  \caption{\textbf{Scattering dynamics with various string tensions.} 
  \textbf{a}, Space-time evolution of the particle density $\rho_i(t)$ for initial two meson-like excitations embedded in a bare vacuum background at different string tensions, controlled by the topological angle term. From left to right, $\chi / \kappa = 0$, 0.3, 0.4, and 0.6.
  \textbf{b}, Corresponding evolution of the dynamical charge $Q_i(t)$. The color bars indicate the value of $\rho$ (\textbf{a}) and $Q$ (\textbf{b}). As the string tension increases, more localized particle dynamics is observed, indicating stronger confinement.
  \textbf{c}, Time evolution of the average electric field $E(t) = [E_9(t) + E_{17}(t)]/2$. In the confinement regime, longer-lived oscillatory dynamics and slower decay rates are observed with increasing string tension. In the deconfinement regime, a rapid decay is observed at early times due to the free propagation of the particles, followed by small residual oscillations tagged by post-collision scattering dynamics during the evolution.
}
  \label{EDFig:string tension}
\end{figure*}

Extended Data Fig.~\ref{EDFig:string tension}\textbf{a--b} show the scattering dynamics via the space-time evolution of the particle density $\rho_i(t)$ (\textbf{a}) and the dynamical charge $Q_i(t)$ (\textbf{b}). At large string tension, e.g., $\chi = 0.6 \kappa$, the initially excited particle–antiparticle pairs remain tightly localized at their initial matter sites, forming stable bound states due to the strong effective string tension, as discussed in the main text. As the string tension is reduced, the particles remain bonded but exhibit wave-packet spreading over extended evolution times. When the string tension is tuned to zero, the bound states fully dissolve, and the particles exhibit scattering dynamics, as described in the main text.

We further probe scattering dynamics by tracing the average electric field initially encoded as negative electric field in the positive vacuum background, given by $E = (E_9 + E_{17})/2$. The experimental results are shown in Extended Data Fig.~\ref{EDFig:string tension}\textbf{c}. In the confinement regime, we observe longer-lived oscillatory dynamics and slower decay rates as the string tension increases, indicating stronger particle localization. In contrast, in the deconfinement regime, we observe a rapid decay of the electric field at early times ($\Omega t \lesssim 12$) due to the free propagation of the particles. Following the collisions of the particles, further scattering dynamics result in small residual oscillations of the electric field, rather than the sustained decaying behavior of the confinement regime.

\vspace{.5cm}
\noindent\textbf{Scattering dynamics with varying masses}

\begin{figure*}[h]
  \includegraphics[width=\textwidth]{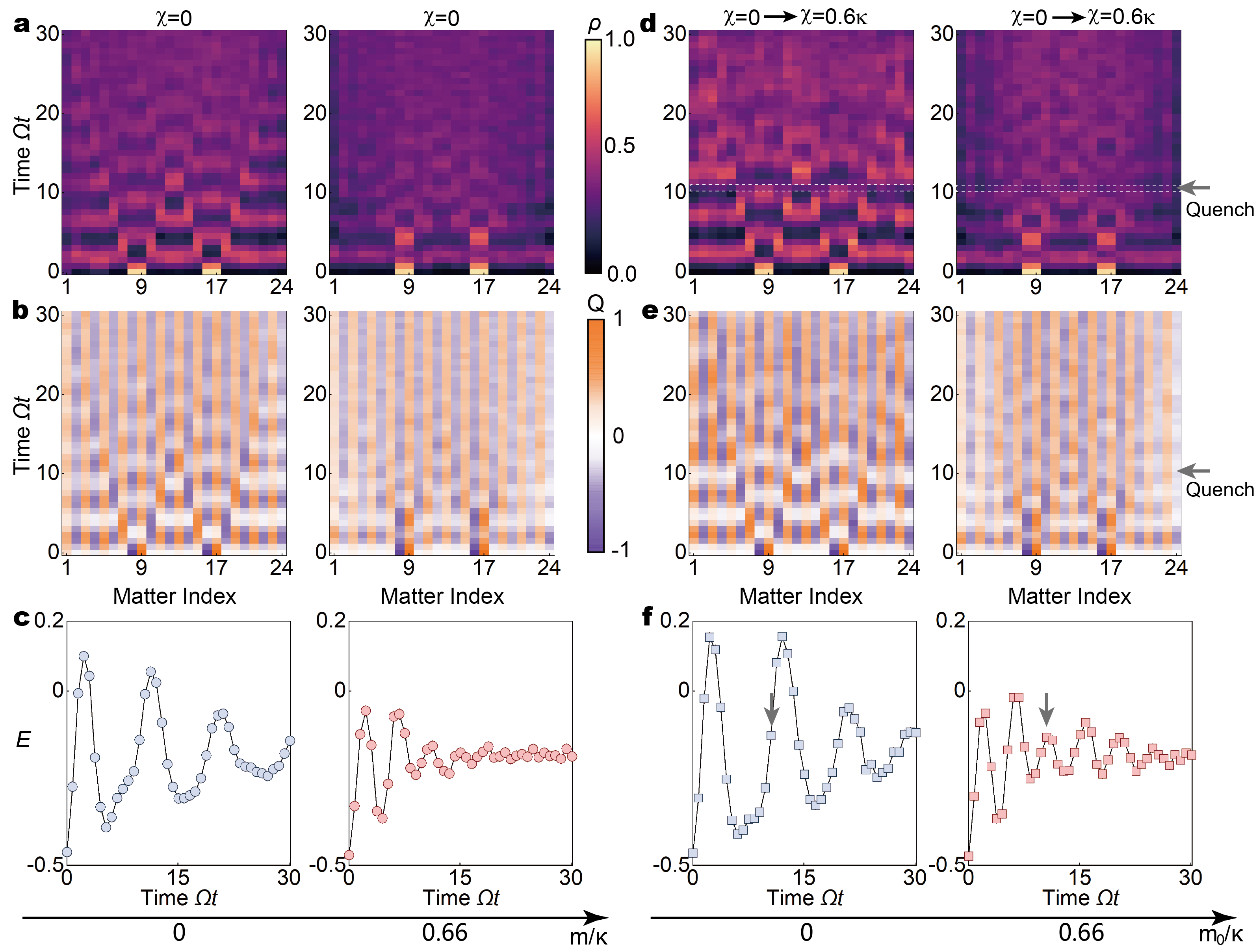}
  \caption{\textbf{Scattering dynamics at tunable fermion masses and string tension quench.} 
  \textbf{a–c}, Dynamics in the deconfinement phase. \textbf{a,b}, Space-time evolution of the particle density $\rho_i(t)$ (\textbf{a}) and dynamical charge $Q_i(t)$ (\textbf{b}) for $m = 0$ (left) and $m = 0.66 \kappa$ (right). For $m=0$, free particle propagation and elastic scattering trajectories are visible, accompanied by pronounced vacuum fluctuations. For $m = 0.66 \kappa$, the dynamics are more complex at the critical mass point. \textbf{c}, Corresponding time evolution of the average electric field $E(t) = [E_9(t) + E_{17}(t)]/2$.  
  \textbf{d–f}, String tension quench dynamics. The initially deconfinement phase is quenched to a confinement phase with $\chi = 0.6 \kappa$ at $\Omega t = 10.6$ (indicated by gray arrows and dashed lines). \textbf{d,e}, Space-time evolution of $\rho_i(t)$ (\textbf{d}) and $Q_i(t)$ (\textbf{e}) for initial $m_0 = 0$ (left) and $m_0 = 0.66 \kappa$ (right). After the quench, elastic scattering trajectories in the interference region disappear for $m_0 = 0$, while pronounced post-quench oscillations emerge for $m_0 = 0.66 \kappa$. \textbf{f}, Evolution of $E(t)$ showing enhanced string oscillations for $m_0 = 0.66 \kappa$ compared to the weak variation for $m_0 = 0$.
}
  \label{EDFig:tune masses}
\end{figure*}

\noindent The dynamics can vary with the fermion mass in both confinement and deconfinement regimes, due to the mass-dependent vacuum fluctuations and the string breaking threshold of $U(L) \sim 2m$. To further investigate these effects, we experimentally probe scattering dynamics at smaller fermion masses. After preparing the meson-like states, we first set the string tension to zero, initializing the system in the deconfinement phase. The dynamics are characterized by the space-time evolution of the particle density $\rho_i(t)$ and dynamical charge $Q_i(t)$, as shown in Extended Data Fig.~\ref{EDFig:tune masses}\textbf{a--b}. For $m=0$, the free propagation of particles is observed, accompanied by particle production and pronounced background vacuum fluctuations. In the central region, clear trajectories arising from particle collisions are visible. When the mass is tuned to the critical value $m_c = 0.66\kappa$, the system is at the transition point where the scalar potential $V(\phi)$ changes from a single-well to a double-well structure. The flatness at the bottom of the potential gives rise to more complex particle dynamics, as evident in Extended Data Fig.~\ref{EDFig:tune masses}\textbf{a--b}.

We then apply a quench protocol to explore non-equilibrium dynamics during scattering. The system is initialized in the deconfinement phase with $m_0 = 0$ or $0.66 \kappa$, allowing for initial free propagation. At $\Omega t = 10.6$ (indicated by the gray arrows in Extended Data Fig.~\ref{EDFig:tune masses}\textbf{d--f}), we suddenly apply local addressing beams (see the Experimental details section), inducing an AC-Stark shift of $\Delta_{\rm ac} = 0.6 \Omega$ on the addressed sites. This effectively introduces a strong string tension of $\chi = 0.6 \kappa$, while the mass is simultaneously shifted to $m_0 + 0.3 \kappa$. The system is thus quenched into the confinement phase.

As shown in Extended Data Fig.~\ref{EDFig:tune masses}\textbf{d--e}, after the quench, the elastic scattering trajectories observed in Extended Data Fig.~\ref{EDFig:tune masses}\textbf{a--b} disappear in the $m_0 = 0$ case, manifesting the confinement effect on particle propagation. For $m_0 = 0.66 \kappa$, stronger electric field oscillatory dynamics are observed, as demonstrated in Extended Data Fig.~\ref{EDFig:tune masses}\textbf{c,f}, where we show the evolution of the average electric field $E(t) = [E_9(t) + E_{17}(t)]/2$. For $m_0 = 0$, the dynamics of $E(t)$ remain relatively unchanged, as the small mass results in a weak confinement effect, consistent with the discussion in the main text. In contrast, for $m_0 = 0.66 \kappa$, pronounced post-quench oscillations reflect the altered vacuum structure. It is noteworthy that no addressing beams are applied to gauge sites 9 and 17 before or after the dual quench. These results demonstrate that, although particle propagation is frozen by the strong string tension after the quench, vacuum oscillations persist, with lighter masses exhibiting more pronounced vacuum fluctuations and heavier masses enabling cleaner dynamics of the initially encoded meson-like excitations, as displayed in the main text.

\vspace{.5cm}
\noindent\textbf{Numerical simulations}

\noindent We simulate the nonequilibrium dynamics of a 1D Rydberg atom chain using matrix product state (MPS) methods implemented via the TeNPy library~\cite{Hauschild2018Efficient}. The system is described by a time-dependent Hamiltonian:
\begin{equation}
\hat{H}(t) = \frac{\Omega}{2}\sum_i\hat{\sigma}_x^i + \sum_i\frac{\Delta_i(t)}{2}\hat{\sigma}_z^i + \sum_{i<j}V_{i, j}\hat{n}_i\hat{n}_{j},   
\end{equation}
where all parameters are experimentally calibrated: Rabi frequency $\Omega$, van der Waals interactions $V_{i,j}$, and $\Delta_i(t)$ including both global and local detunings. In all numerical simulations, we focus on the dominant nearest-neighbor and next-nearest-neighbor interactions while neglecting the weaker longer-range terms for computational efficiency. Time evolution following a quench at $t = t_{\text{quench}}$ is computed using the two-site time-dependent variational principle (TDVP) with a fixed time step $\Omega\Delta t = 0.06\pi$, chosen to balance numerical stability and accuracy. The MPS simulations employ a bond dimension $\chi \leq 100$ with singular value cutoff $10^{-8}$, providing an optimal trade-off between computational efficiency and precision.

In TeNPy-based numerical simulations that map the experimental multi-level system to an effective two-level system, we incorporate two primary sources of error to model experimental conditions: state preparation imperfections and the combined effects of depolarization (bit-flip type) and detection errors.

\noindent \textbf{State preparation errors.} To quantify the impact of imperfect state preparation (detailed in the error analysis section in Method), and based on the measured high fidelity of the initial state preparation, we perform $L+1$ parallel simulations (where $L$ is the length of the chain): one with the ideal initial state $\psi_{\text{ideal}}$ and $L$ additional simulations, each containing a single spin-flip error $\ket{\psi_i} = \hat{\sigma}_x^i\ket{\psi_{\text{ideal}}}$ at site $i$. It should be noted that this approximation may become less effective as the preparation efficiency decreases further, since the microscopic error distribution could shift toward more complex multi-qubit error patterns. Using the experimentally measured preparation fidelity $\mathcal{F}$, we construct a weighted ensemble:
\begin{equation}
\rho_{\text{exp}} = \mathcal{F}\ket{\psi_{\text{ideal}}}\bra{\psi_{\text{ideal}}} + \frac{1-\mathcal{F}}{L}\sum_{i=1}^L\ket{\psi_i}\bra{\psi_i}.
\end{equation}

\noindent \textbf{Depolarization and detection errors.} The experimental system experiences dissipative processes during time evolution due to intermediate state spontaneous emission and finite Rydberg lifetime. However, the TenPy package cannot directly incorporate these dissipative processes for our large 29-atom chain due to the limitations of matrix product state methods for simulating open quantum dynamics. Here, we model these dissipative processes---which primarily cause atoms originally in Rydberg states to flip to ground states as bit-flip type depolarization---as equivalent detection errors of the Rydberg states, due to limited computational resources. This model is justified by several key observations: First, both depolarization and detection errors manifest as discrepancies between the actual and measured quantum states. Second, for the observables we measure, the dominant effect of Rydberg flip-errors contributes to misclassifying Rydberg states as ground states after evolution concludes. The additional Rydberg flip-errors can be understood by considering the two dominant loss mechanisms: During the Hamiltonian driving (duration $\sim 4\,\mu$s in the experimental sequence), intermediate state losses contribute $\sim 3.6\%$. Additionally, the finite lifetime of Rydberg states contributes losses throughout the entire experimental sequence from state preparation to final detection (total duration $\sim 5\,\mu$s), accounting for an additional $\sim 3.3\%$. Additionally, the Rydberg state detection error is about 1\% in our experimental setup.

\begin{figure}[h]
  \includegraphics[width=\columnwidth]{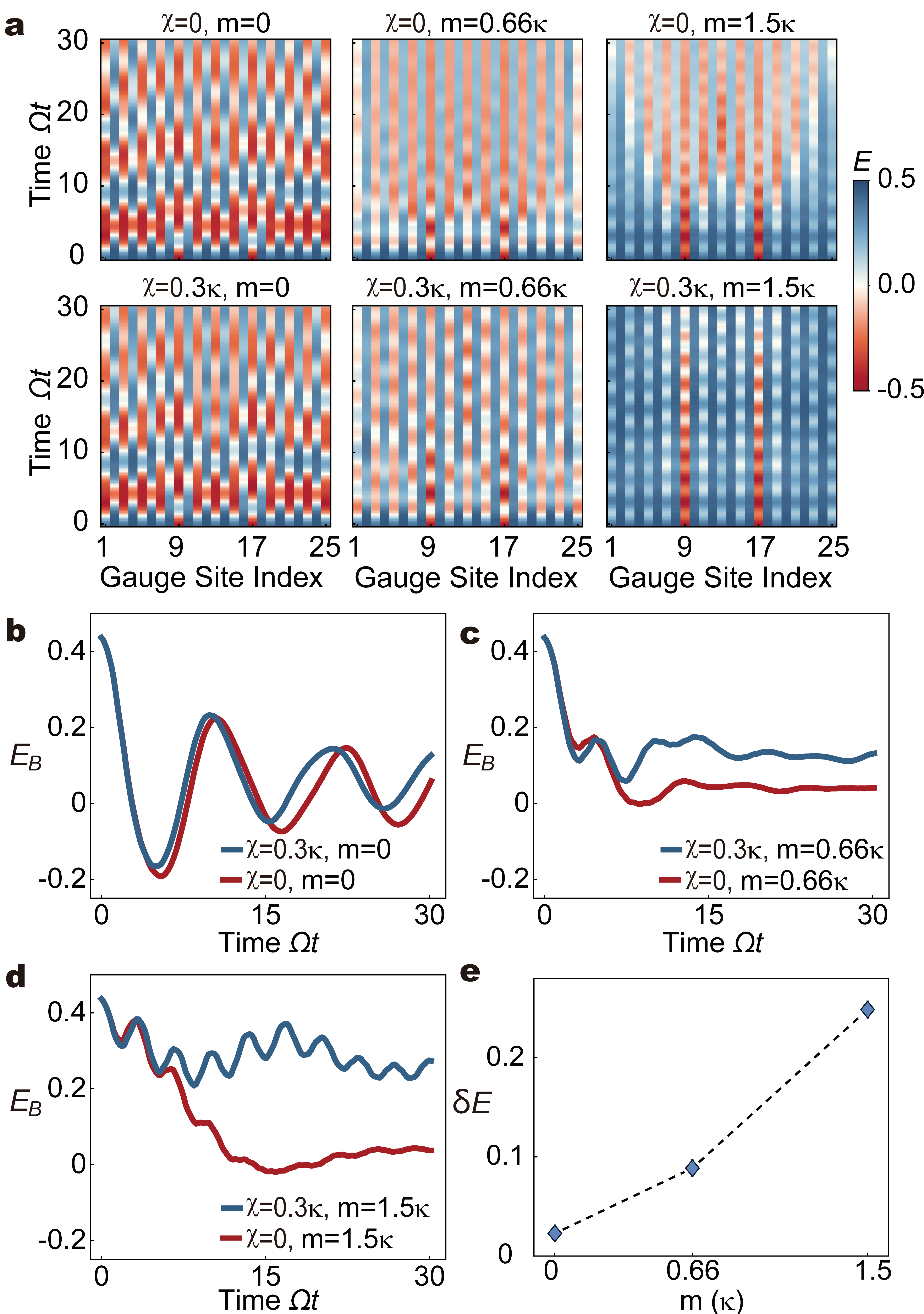}
  \caption{\textbf{Numerical simulations for confinement and deconfinemnt phases.} 
 Simulated spatiotemporal evolution of the electric field $E_i(t)$ (\textbf{a}), bulk electric string $E_B(t)$ (\textbf{b-d}) and steady-state bias $\delta E$ (\textbf{e}) corresponding to the experimental observations shown in Fig.~\ref{Fig:2_confinement_and_deconfinement_phases} of the main text; Great agreement between the experimental data and numerical results is observed.
}
  \label{EDFig:Simulation_2}
\end{figure}

\begin{figure}[h]
  \includegraphics[width=\columnwidth]{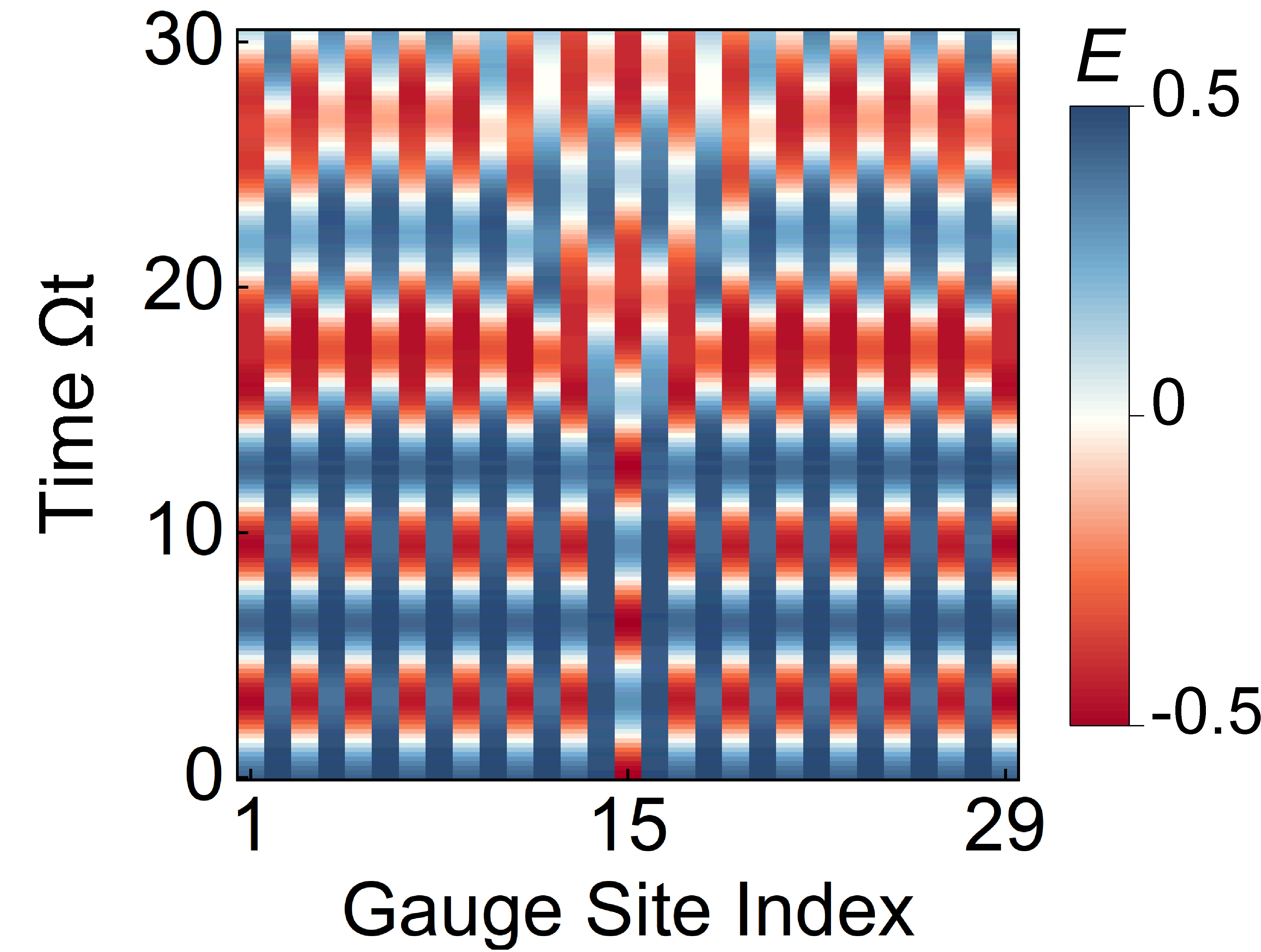}
  \caption{\textbf{Numerical simulations for quench dynamics.} 
  Simulated spatiotemporal evolution of the electric field $E_i(t)$ for the quench dynamics presented in Fig.~\ref{Fig3:Quench_dynamics} of the main text. The simulation captures the rapid string fragmentation following the dual-parameter quench from the confined to deconfined phase at $\Omega^*t=11.3$.
}
  \label{EDFig:Simulation_3}
\end{figure}

\begin{figure}[h]
  \includegraphics[width=\columnwidth]{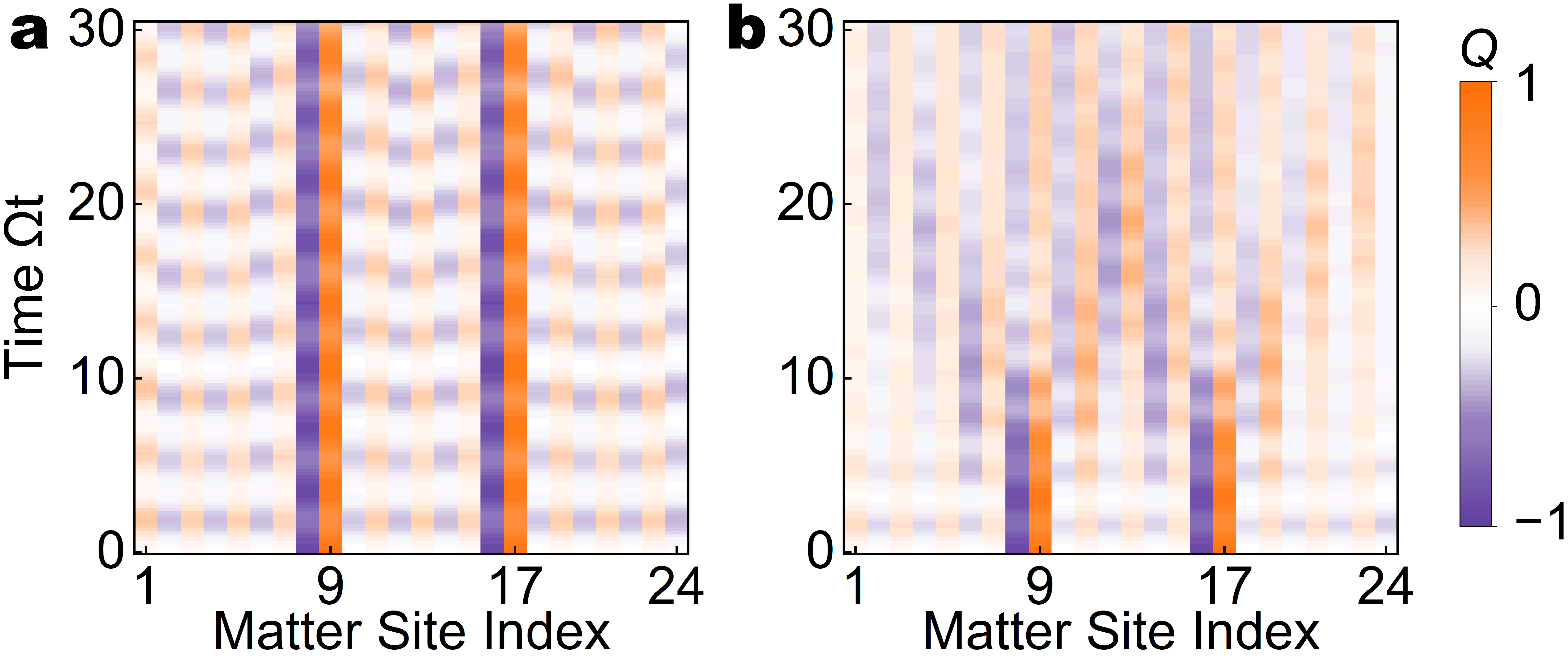}
  \caption{\textbf{Numerical simulations for scattering dynamics.} 
  Simulated spatiotemporal evolution of the dynamical charge $Q_i(t)$ corresponding to the charge scattering dynamics shown in Fig.~\ref{Fig:4_Meson_dynamics}a (confinement phase) and \ref{Fig:4_Meson_dynamics}b (deconfinement phase) of the main text.
}
  \label{EDFig:Simulation_4}
\end{figure}

In these cases, we model the combined depolarization and detection errors by introducing state-dependent detection probabilities: A Rydberg-state atom has a modeled probability $P_r = 8\%$ of being incorrectly detected as a ground-state atom, while a ground-state atom has an experimentally measured probability $P_g = 2\%$ of being misidentified as a Rydberg-state atom. The numerical results of the experimental observables are shown in Extended Data Figs.~\ref{EDFig:Simulation_2},~\ref{EDFig:Simulation_3}, and~\ref{EDFig:Simulation_4}, corresponding to raw experimental observations shown in Figs.~\ref{Fig:2_confinement_and_deconfinement_phases},~\ref{Fig3:Quench_dynamics}, and~\ref{Fig:4_Meson_dynamics} of the main text, respectively. The good agreement between the experimental data and numerical simulations demonstrates the effectiveness of the error model. 
We note that minor deviations still exist between the raw experimental data and the numerical results given by the error model, which are supposed to arise from decoherence caused by other error sources during the Hamiltonian driving that are not considered in our error model, such as laser noise and Doppler effects.

\vspace{.3cm}
\end{document}